\documentclass[12pt, draftclsnofoot, onecolumn]{IEEEtran}
\pdfoutput=1 

\usepackage{color}
\usepackage{soul}
\setulcolor{red}
\sethlcolor{yellow}

\usepackage{caption}
\usepackage{amsmath}
\def\@viiipt{8}
\def\squareforqed{\hbox{\rlap{$\sqcap$}$\sqcup$}}
\def\qed{\ifmmode\squareforqed\else{\unskip\nobreak\hfil
\penalty50\hskip1em\null\nobreak\hfil\squareforqed
\parfillskip=0pt\finalhyphendemerits=0\endgraf}\fi}

\usepackage{graphicx, amssymb, subfigure, setspace}
\usepackage{epstopdf}
\usepackage{bbding}
\usepackage{algorithmicx}
\usepackage{algpseudocode}
\usepackage[ruled]{algorithm}
\usepackage{amsmath}
\usepackage{wrapfig}
\usepackage{textcomp}
\usepackage{url}
\usepackage{textcomp}
\usepackage{bbm}
\usepackage{footnote}

\usepackage{cite}

\usepackage[long]{optional}

\usepackage{comment}
\usepackage{bm}

\usepackage{relsize}

\usepackage{amsmath}
\allowdisplaybreaks[4]

\begin{document}


%
%
%
%

\title{Online Service Placement and Request Scheduling in MEC Networks }

\author{Lina~Su,
        Ne~Wang,
        ~Ruiting~Zhou,
       and~Zongpeng~Li
\IEEEcompsocitemizethanks{\IEEEcompsocthanksitem L. Su, Z. Li and N. Wang are with the School of Computer Science, Wuhan University, Wuhan 430072, China.
E-mail: \{lina.su, ne.wang, zongpeng\}@whu.edu.cn.
\IEEEcompsocthanksitem R. Zhou is with the Key Laboratory of Aerospace Information Security and Trusted Computing, Ministry of Education, School of Cyber Science and Engineering, Wuhan University, Wuhan 430072, China. E-mail: ruitingzhou@whu.edu.cn.}
\thanks{Manuscript received August 18, 2021.(Corresponding author: Zongpeng Li.)}}

\maketitle
\begin{abstract}
Mobile edge computing (MEC) emerges as a promising solution for servicing delay-sensitive tasks at the edge network. A body of recent literature started to focus on cost-efficient service placement and request scheduling. This work investigates the joint optimization of service placement and request scheduling in a dense MEC network, and develops an efficient online algorithm that achieves close-to-optimal performance. Our online algorithm consists of two basic modules: (1) a regularization with look-ahead approach from competitive online convex optimization, for decomposing the offline relaxed minimization problem into multiple sub-problems, each of which can be efficiently solved in each time slot; (2) a randomized rounding method to transform the fractional solution of offline relaxed problem into integer solution of the original minimization problem, guaranteeing a low competitive ratio. Both theoretical analysis and simulation studies corroborate the efficacy of our proposed online MEC optimization algorithm.

\end{abstract}


%
%
%

%
%

%
%


\section{Introduction}
\subsection{Background and Motivations}
\IEEEPARstart{M}{any} intelligent applications including self-driving, augment reality, and virtual reality group gaming have spurred a increasing demand of low-latency services that may not be solely satisfied by today's centralized cloud architecture. MEC~\cite{MEC,MEC-2} emerges as a key technique to accommodate the need by pushing substantial amounts of computational capabilities to the network edge, in the vicinity of mobile users.

With MEC, services can be hosted at various types of small-cell base stations (SBSs), endowed with storage resources for processing service requests. User requests for these services can be completed by the local SBSs without duplicate transmission from a central server, helping improve quality of service and reduce operational cost. Due to its limited capacity, each SBS does not provide satisfactory services to all users. Therefore, this leads to the question of which services to be hosted and where to execute each service in order to achieve economic efficiency and meanwhile offer high-quality services.

Fig.1 illustrates an example with five heterogeneous SBSs. Initially, each SBS places a subset of services. If service {\small$k_1$} is already hosted in SBS {\small$S_1$}, user {\small$u_1$} requesting service {\small$k_1$} can be served locally. Otherwise, user may be served by non-nearest SBSs at the cost of additional communication delay (as user {\small$u_2$} illustrated). With latest 5G technology, SBSs can communicate with others beyond working individually. Furthermore, to better meet subsequent requests, each SBS has to dynamically manage its services by removing or replacing the current services (as SBS {\small$S_3,S_4,S_5$} illustrated).

\vspace{-6mm}
\begin{figure}[htbp] 
	\centering
	\includegraphics[width=0.5\textwidth]{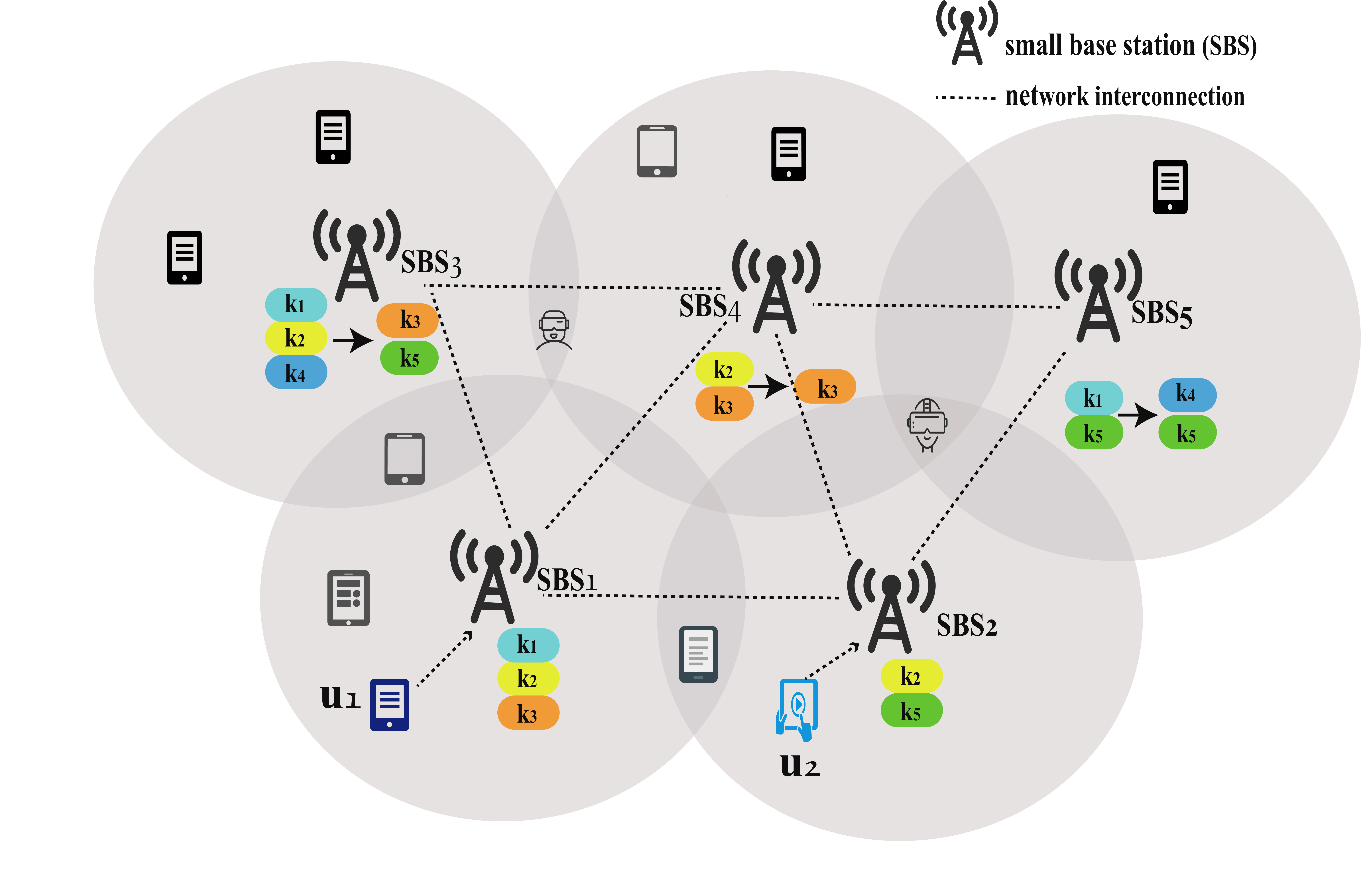}	
	\vspace{-1mm}
	\caption{An illustration of the MEC system.}
\end{figure}
\vspace{-4mm}

In addition to storage resources, modern innovative services also require data transmission, which takes up a proportion of SBS bandwidth capacities and may even cause network congestion. Moreover, under dense deployment of SBSs in 5G network~\cite{5G}, mobile users often reside in the overlap coverage of SBSs. For the operator, bandwidth requirement and complicated multi-SBSs scenario make placing and scheduling strategy more challenging. In this context, system operator has a dynamic repertoire of service placing and scheduling alternatives. To minimize overall operational cost, the operator should jointly optimize these strategies in an online fashion.

Previous studies examined the joint placing and routing problem for different objectives, including minimizing service access latency~\cite{latency-2,latency-1,latency-3,Xu,2019Winning,XuJ} and maximizing the volume of served requests~\cite{KP,cache-hit-1,KP-2}. A few recent  approaches focus on cost-efficient placing and scheduling. They suffer from limiting assumptions and lack of performance guarantee. Heuristic algorithms without provable performance guarantee are proposed in\cite{Taleb,Yang,Ceselli}. Among those algorithms with theoretical approximation guarantees~\cite{2020Intelligent,Zeng,Zhao}, the performances are usually unsatisfactory.

Taking the above issue into account, the key challenges are:

\emph{How to dynamically adjust service placement in SBSs to satisfy the time-varying requests}?

\emph{How to timely schedule user requests to appropriate SBSs}?

\emph{How to jointly optimize these strategies in an online manner to minimize holistic operational cost}?

\subsection{Methodology and Contributions}

In this work, we establish a comprehensive model to address the above challenges, summarized as follows.

{\small$\bullet$} \emph{Online Framework}. We formulate the joint placement and scheduling problem in MEC networks, aiming to minimize holistic operational cost. We consider an MEC system with 1) heterogeneous SBSs and diverse services, 2) stringent capacity limitations of SBSs, 3) time-varying user requests, and 4) overlapping coverage regions of multi-SBSs.

{\small$\bullet$} \emph{Regularization with look-Ahead Algorithm}. We employ a regularization with look-ahead approach from competitive online convex optimization (OCO)~\cite{OCO} to translate the offline relaxed problem into a more tractable problem. The latter is then decomposed into multiple versions, each of which partitions the time horizon into multiple episodes. Solving the problem of each episode in each version, we obtain feasible fractional solutions to the offline relaxed problem by applying Karush-Kuhn-Tucker (KKT) optimality conditions, while rigorously proving an upper-bound on holistic operational cost relative to offline optimum.

{\small$\bullet$} \emph{Rounding Algorithm}. We devise a novel randomized rounding algorithm to transform fractional solution of the offline relaxed problem into integer ones of the original problem. Working in concert with the regularization with look-ahead algorithm, a low competitive ratio can be realized.

{\small$\bullet$} \emph{Evaluation Results}. We carry out simulations in order to evaluate the effectiveness of the proposed online framework. Compared with the latest related schemes, our proposed online framework has a superior performance.
 \section{Literature Review}\label{sec:related}
The problem of MEC service placement and scheduling has received increased attention in recent years. 
The majority of previous works has been devoted to minimizing the latency~\cite{latency-2,latency-1,latency-3,Xu,2019Winning,XuJ} or maximizing the volume of served requests~\cite{KP,cache-hit-1,KP-2}. 

Tran et al.~\cite{latency-2} study how to minimize video access latency by leveraging the conditional gradient method. 
Dehghan et al.~\cite{latency-1} apply the greedy strategy and submodular optimization to develop an approximation algorithm.
Li et al.~\cite{latency-3} present a distributed resilient caching algorithm according to the concave relaxation of expected content access latency.
Xu et al.~\cite{Xu} define the joint problem as an integer linear programming (ILP) problem and develop a heuristic algorithm.
The works of~\cite{2019Winning,XuJ} both exploit Lyapunov optimization approach to study the time-average latency, given diverse services and demands, and decentralized cooperation.
Poularakis et al.~\cite{KP} investigate the joint optimization problem in dense MEC networks, and formulate it as a submodular optimization problem respecting triple practical constraints.
Considering the sharable (storage) and non-sharable (computation, link bandwidth) resources, He et al.~\cite{cache-hit-1} assume non-overlapped coverage region among multiple base stations and propose a greedy algorithm.
 
While a few recent interesting studies focus on cost-efficient placing and scheduling, most of them still has certain restrictions in the light of practicality and performance guarantee. 

Taleb et al.~\cite{Taleb} advocate content delivery network slicing, formulate cost optimization into an ILP, and design computational-efficient heuristics.
Yang et al.~\cite{Yang} utilize online request prediction to regulate each round solution close to the optimum by a greedy algorithm. 
Ceselli et al.~\cite{Ceselli} study the minimization of installation costs in network facilities, and design an heuristic algorithm.
The works of~\cite{Taleb,Yang,Ceselli} fail to provide theoretical approximation guarantee.
Deep reinforcement learning has been adopted in \cite{2020Intelligent}, while their model confines service to identically sized content, and only minimizes the traffic cost.
Zeng et al.~\cite{Zeng} utilize primal-dual decomposition to translate the joint problem into bi-problems, and design an approximation algorithm to solve each sub-problem.
Zhao et al.~\cite{Zhao} highlight the cost of forwarding requests and downloading services in a homogeneous edge-cloud setting where all services consume the same amount of storage capacity and only resides in one edge node. However, those algorithms~\cite{2020Intelligent,Zeng,Zhao} provide unsatisfactory competitive ratios.

Departing from the above literature, we aim to develop a more efficient online algorithm with low competitive ratio, through a meticulously arranged marriage of competitive OCO and ILP rounding.

\section{System Model and Problem Definition}\label{sec:model}
\subsection{The MEC network}\label{sec:sub_The_MEC_network}

The structure of MEC network, as depicted in Fig.1, consists of a set of geo-distributed SBSs, and a set of mobile users (MUs). Let  {\small$\mathcal{M}=\{1,2,\dots,M\}$} and {\small$\mathcal{N}=\{1,2,\dots,N\}$} represent the SBS set and MU set, respectively. The time horizon is discretized into multiple time-frames, indexed by {\small$t \in \{1,2,\dots,T\}$}. The duration {\small $T$} may be three days and a time slot {\small$t$} may be five minutes, respectively.

Considering the heterogeneity of SBSs, we assume that SBS {\small $m$} has a storage capacity {\small $R_m$} to pre-store services and a bandwidth capacity {\small $C_m$} to ensure service transmission. The MEC system supports a library of delay-sensitive or data-intensive services, denoted by {\small$\mathcal{K}=\{1,2,\dots,K\}$}. Different services as exemplified by multi-media content caching and virtual reality group gaming may occupy distinct resource capacities. Let {\small $r_k$} and {\small $c_k$} indicate the storage and bandwidth resource occupied by service {\small $k$}.

The service requests arrive randomly. The amount of requests from user {\small$n$} for service {\small $k$} in {\small $t$} is denoted as {$\lambda_{n,k}(t)$}. Each user can get served by the nearest or non-nearest SBSs, depending on the joint placing and scheduling strategies of MEC operator. Provided that the requested service is locally housed and the SBS has adequate bandwidth resource, the request can be dispatched to the nearest SBS. If the nearest SBS has not stored or sufficient bandwidth, it resorts to its connective SBSs while inevitably generating additional communication cost.

Due to the finite capacities, the MEC operator has to timely determine which services to place in SBSs and how to schedule real-time requests to the SBSs. Thus, we introduce two set of optimization variables:
(i) service placing decision {\small $x_{m,k}(t) \in \{0,1\}$}, which indicates whether SBS {\small $m$} stores a replica of service {\small $k$} in {\small $t$} or not; 
(ii) request scheduling decision {\small $y_{m,n,k}(t) \in \left[ 0,1 \right]$}, which implies SBS {\small $m$} serves the proportion of {$\lambda_{n,k}(t)$} requests.

\subsection{Cost Structure}\label{sec:sub_cost_function}
The holistic cost of MEC system in its running time consists of three components.

{\bf\em Storage cost} for storing service. The storage cost in {\small$t$} is computed as:

\vspace{-6mm}
{\small
\begin{equation}
	C_R(t)=\sum_{m}\sum_{k}l_{m,k}{x_{m,k}(t)},
\end{equation}}
\vspace{-6mm}

\noindent where {\small $l_{m,k}$} denote the unit storage cost to store service {\small$k$} in SBS {\small$m$}.

{\bf\em Service cost of SBSs.} For each SBS, service cost mainly depends on MUs' relative locations, the number of served requests, and the proportion of services decided by the scheduling strategy. In each time frame, let {\small $d_{m,n}>0$} describe the communication parameter to weigh the location of MUs. For example, when MUs locate in the boundary of the SBSs, serving such MUs consume more communication power, which leads to higher cost. Accordingly, the service cost of SBSs in {\small$t$} is:

\vspace{-6mm}
{\small
\begin{equation}
	C_S(t)=\sum_{m}\sum_{n}\sum_{k}d_{m,n}{\lambda_{n,k}(t)}{y_{m,n,k}(t)}.
\end{equation}}
\vspace{-6mm}

{\bf\em Dynamic service placement cost.} In addition to dispatch the real-time requests, the MEC system has to dynamically manage the service placing of each SBS by removing current services or placing newly service. To this end, we introduce a binary variable {\small $z_{m,k}(t)$} to represent whether SBS {\small $m$} stores newly service {\small $k$} in {\small$t$} or not, {\em i.e.,} {\small${z_{m,k}(t)}= max \{x_{m,k}(t)-x_{m,k}(t-1),0\}$}. The dynamic service placing cost in {\small $t$} is

\vspace{-6mm}
{\small
\begin{equation}
	C_D(t)=\sum_{m}\sum_{k}b_{m,k}{z_{m,k}(t)},
\end{equation}}
\vspace{-6mm}

\noindent where {\small $b_{m,k}$} denote the unit placing cost of service {\small$k$} in SBS {\small$m$}.

\subsection{Problem definition}
We investigate the offline setting where all user requests are absolutely known, aiming at minimizing the holistic cost. By adding Eq(1) to Eq(3) together, the offline problem, denoted by {\small$Cost$}, can be formulated as follows

\vspace{-3mm}
{\small
\begin{equation}
\label{eq:min}
\mbox{\\  minimize  }
	\sum_{t}\bigg(C_R(t)+C_S(t)+C_D(t)\bigg)
\end{equation}

subject to:
\begin{align}
 y_{m,n,k}(t)\leq x_{m,k}(t), \forall m,n,k,t,
 \tag{\ref{eq:min}a}\\
\sum_{m}y_{m,n,k}(t)\geq 1,\forall n,k,t, \tag{\ref{eq:min}b}\\
	z_{m,k}(t) \geq x_{m,k}(t)-x_{m,k}(t-1),\forall m,k,t, \tag{\ref{eq:min}c }\\
\sum_{k}x_{m,k}(t)r_k\leq R_m,\forall m,t, \tag{\ref{eq:min}d}\\
\sum_{n}\sum_{k}y_{m,n,k}(t)\lambda_{n,k}(t)c_k \leq C_m,\forall m,t, \tag{\ref{eq:min}e}\\
y_{m,n,k}(t)\in \left[ 0,1 \right],\forall m,n,k,t, \tag{\ref{eq:min}f}\\
x_{m,k}(t) \in\{ 0,1 \},\forall m,k,t, \tag{\ref{eq:min}g}\\
z_{m,k}(t) \in\{ 0,1 \},\forall m,k,t, \tag{\ref{eq:min}h}
\end{align}}
\vspace{-1mm}
\noindent where {\small$\forall m,n,k,t$} represent
{\small$\forall m\in\mathcal{M},n\in\mathcal{N}, k\in\mathcal{K}, t\in\mathcal{T}$}.

Constraint (\ref{eq:min}a) shows that a SBS should have already stored a replica of the requested service before handling a request. Constraint (\ref{eq:min}b) indicates that the user request can be assigned to one or multiple SBSs. Constraint (\ref{eq:min}d) and (\ref{eq:min}e) imply demand-supply balance conditions that SBS {\small $m$} provided service can not exceed its capacities regarding storage and bandwidth resource.

{\bf\em Algorithmic challenges.} The main challenges of minimization problem are two-fold. Firstly, in realistic MEC system, it is not readily to make wise decisions timely due to the system dynamics, such as newly arrived requests and placed services. Secondly, even within the offline scenario where all the external inputs are absolutely provided, the original problem is still a mixed-integer linear programming, which is NP-hard~\cite{1980Complexity}. Then, without knowing all the future information, how to online optimize holistic operational cost.

\subsection{Basic idea}

Enlightened by the competitive OCO~\cite{OCO}, we devise an efficient online algorithm that leverages the regularization with look-ahead approach and rounding technique, to tackle the above challenges. The algorithmic idea, as illustrated in Fig.2, can be partitioned into two steps.

{\bf\em Step 1.} In Sec.~\ref{sec:ORA}, we first obtain the fractional minimization problem {\small $P$} via relaxing the integer decision variables. Then, by regularization with look-ahead, the fractional problem {\small $P$} can be transformed into a more tractable problem {\small $P_{ORA}$}. The new problem {\small $P_{ORA}$} can be readily decomposed into multiple versions, each of which partitions the time domain into a series of episodes. And then, the dynamic service placing overhead of the beginning and last time-frame in individual episode are substituted by two carefully-modified regularized terms in the objective function. Solving the problem of each episode in each version, the fractional solution to relaxed minimization problem is derived through online algorithm {\small $ORA$}.

{\bf\em Step 2.} In Sec.~\ref{sec:rounding_algorithm}, considering the integral feature of decision variables, we design a randomized dependent service placement algorithm {\small $RDSP$}, so as to round the fractional solution derived by {\small $ ORA$} into integer ones for the original problem.

Next, we discuss our online algorithm in details. 

\vspace{-3mm}
\begin{figure}[htbp] 
	\centering
	\includegraphics[width=0.8\textwidth]{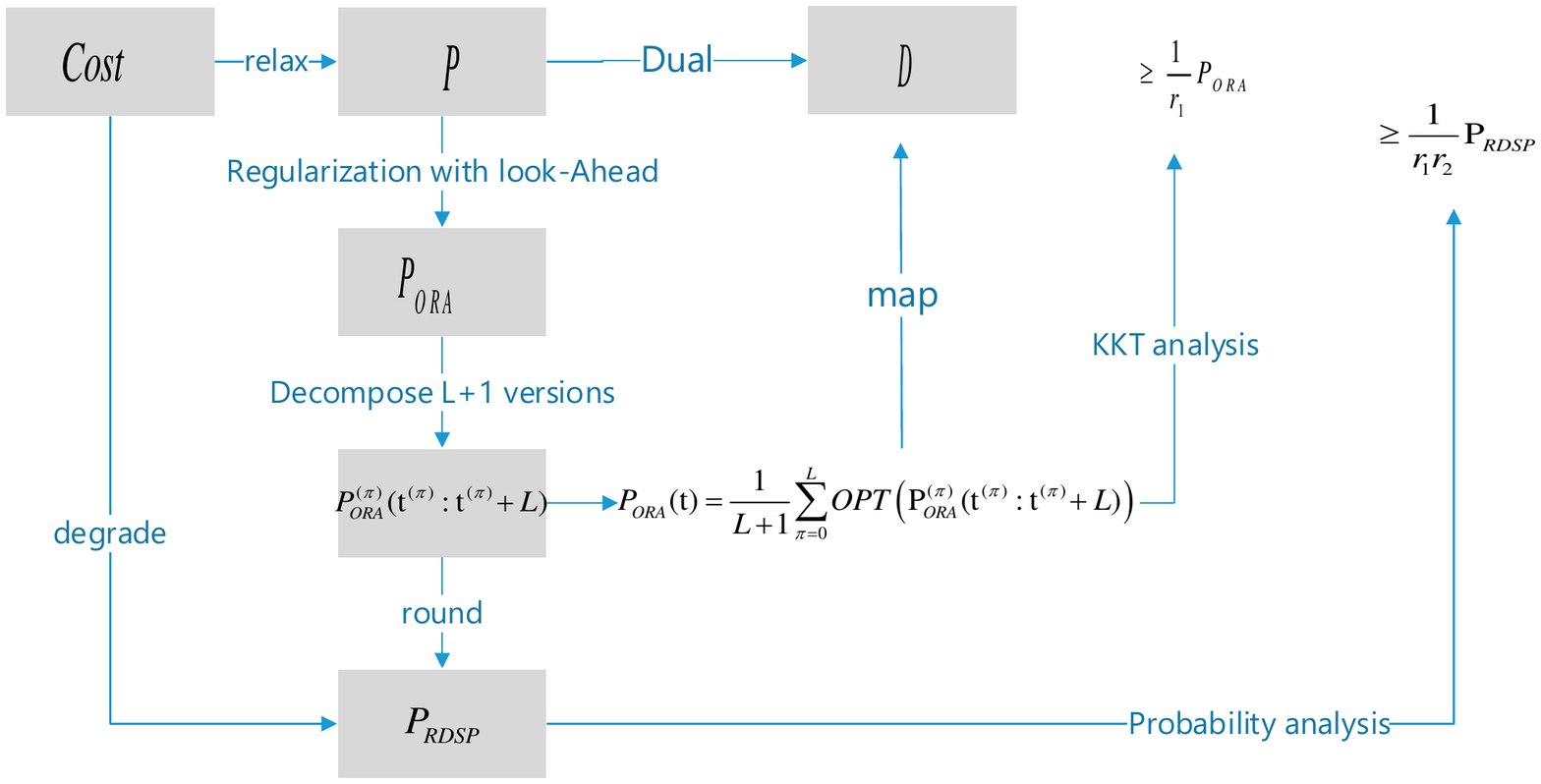}
	\vspace{-6mm}
	\caption{An illustration of our online algorithm}
\end{figure}

\section{Online Regularization with Look-Ahead Algorithm }\label{sec:ORA}

\subsection{Online Regularization with Look-Ahead Algorithm}\label{sec:sub_Problem_ Decomposition}
1) Motivation

To conquer the challenge of NP-hardness in the offline minimization problem, we first relax the integral variables, deriving the fractional minimization problem {\small $P$} as follows:

\vspace{-3mm}
{\small
\begin{equation}
\label{eq:min2}
\mbox{\\ minimize  }
	\sum_{t}\bigg(C_R(t)+C_S(t)+C_D(t)\bigg)
\end{equation}	

subject to:  
  (\ref{eq:min}a) - (\ref{eq:min}f),
\begin{align}
 x_{m,k}(t) \in \left[ 0,1 \right],\forall m,k,t, \tag{\ref{eq:min2}g} \\
  z_{m,k}(t) \in \left[ 0,1 \right],\forall m,k,t. \tag{\ref{eq:min2}h}
\end{align}}
\vspace{-6mm}

For the fractional minimization problem {\small $P$}, a straightforward solution is to greedily employ the best control solutions in each time frame. However, such simple scheme would not certainly guarantee the global optimal solution for the time span, and may even result in bad results~\cite{zhouzhi}.

Towards the competitive performance guarantee, we leverage the regularization technique and finite look-ahead information to design a novel online algorithm, called Online Regularization with look-Ahead (ORA). Here, look-ahead indicates that the system operator knows the current information, and the information of consecutively following time (i.e., the look-ahead window size).  The key idea of algorithm ORA is to adopt two carefully-designed regularization terms to approximate the dynamic service placing overhead of the beginning and last time-frame in 
every episode.

More specifically, we first denote {\small $\pi$} as a non-negative integer from {\small $0$} to {\small $L$}. The algorithm ORA executes {\small $L+1$} versions. Let {\small $ORA^{(\pi)}$} represent the {\small $\pi$}-th version of ORA. {\small$ORA^{(\pi)}$} partitions the entire time into multiple episodes, each of which varies from {\small $t^{(\pi)}$} to {\small $t^{(\pi)}+L$}. And {\small $t^{(\pi)}=\pi+(L+1)v
$}, where {\small $v=-1,0,\dots,\lceil \frac{T}{L+1} \rceil$}. In time {\small $t^{(\pi)}$}, the current information and the consecutively following {\small $L$} time-slots' information are provided. Thus, we further derive a formulation of {\small $ORA^{(\pi)}$}, denoted by {\small $P_{ORA}^{(\pi)}$}:

\vspace{-6mm}
{\small
\begin{equation}
\label{eq:min3}
\begin{aligned}
\text{minimize  }	
  & \sum\limits_{s=t^{(\pi)}}^{t^{(\pi)}+L}\sum\limits_{m}\sum\limits_{k}l_{m,k}x_{m,k}(s)\\
  &+\sum\limits_{s=t^{(\pi)}}^{t^{(\pi)}+L}\sum\limits_{m}\sum\limits_{n}\sum\limits_{k}d_{m,n}{\lambda_{n,k}(s)}{y_{m,n,k}(s)}\\
  & +\sum\limits_{m}\sum\limits_{k}\frac{b_{m,k}}{\eta}x_{m,k}(t^{(\pi)})\ln\left(\frac{1+\frac{\varepsilon}{\mathcal{M}\mathcal{K}}}{x_{m,k}^{(\pi)}(t^{(\pi)}-1)+\frac{\varepsilon}{\mathcal{M}\mathcal{K}}}\right)\\
  &+\sum\limits_{s=t^{(\pi)}+1}^{t^{(\pi)}+L}\sum\limits_{m}\sum\limits_{k}b_{m,k}z_{m,k}(s)\\
  &+\sum\limits_{m}\sum\limits_{k}\frac{b_{m,k}}{\eta} \left[\left(x_{m,k}(t^{(\pi)}+L)+\frac{\varepsilon}{\mathcal{M}\mathcal{K}}\right)\cdot \ln\left(\frac{x_{m,k}(t^{(\pi)}+L)+\frac{\varepsilon}{\mathcal{M}\mathcal{K}}}{1+\frac{\varepsilon}{\mathcal{M}\mathcal{K}}}\right)-x_{m,k}(t^{(\pi)}+L) \right]  
\end{aligned}
\end{equation}}

{\small
subject to: 
\begin{align}
 y_{m,n,k}(s)\leq x_{m,k}(s), \forall m,n,k,s,
 \tag{\ref{eq:min3}a}\\
\sum_{m}y_{m,n,k}(s)\geq 1,\forall n,k,s, \tag{\ref{eq:min3}b}\\
	z_{m,k}(s) \geq x_{m,k}(s)-x_{m,k}(s-1),\forall m,k,s,\tag{\ref{eq:min3}c }\\
\sum_{k}x_{m,k}(s)r_k\leq R_m,\forall m,s, \tag{\ref{eq:min3}d}\\
\sum_{n}\sum_{k}y_{m,n,k}(s)\lambda_{n,k}(s)c_k\leq C_m,\forall m,s, \tag{\ref{eq:min3}e}\\
x_{m,k}(s), y_{m,n,k}(s), z_{m,k}(s)\in \left[ 0,1 \right],\forall m,n,k,s, \tag{\ref{eq:min3}f}
\end{align}}
\vspace{-1mm}

\noindent where {\small$\forall m,n,k,s$} represent {\small$\forall m\in\mathcal{M},n\in\mathcal{N}, k\in\mathcal{K}$}, and {\small$s\in[t^{(\pi)},t^{(\pi)}+L]$}. The notation {\small$\eta=\ln(1+\frac{\mathcal{M}\mathcal{K}}{\varepsilon})$}, {\small$\varepsilon >0 $} and control decision {\small$x_{m,k}^{(\pi)}(t^{(\pi)}-1)$} can be obtained by solving {\small $P_{ORA}^{(\pi)}$} based on the previous episodes from {\small$t^{(\pi)}-L-1$} to {\small$t^{(\pi)}-1$}.

We should emphasize that, {\small $P_{ORA}^{(\pi)}$} approximates the dynamic service placing overhead of the beginning time-frame {\small$t^{(\pi)}$} and last time-frame {\small$t^{(\pi)}+L$} in present episode by two regularized terms, {\em i.e.,} the third and fifth term in (\ref{eq:min3}), which together guarantee the dynamic service placing overhead at the boundary between the consecutive episodes cannot be arbitrarily large. Owing to the convexity of regularization term, it has been widely adopted to approximate online optimization problems in online learning~\cite{regularization}.

2) Algorithm Design

Since ach problem {\small $P_{ORA}^{(\pi)}$} is a standard convex optimization problem comprising multiple packing and covering constraints, it can be efficiently solved by invoking the interior point method in convex optimization~\cite{Interior}. Then, in each time frame {\small$t$}, ORA gets the mean of {\small $X^{(\pi)}(t)$} and {\small $Y^{(\pi)}(t)$} of all {\small $\pi$} as the control decision of ORA in {\small$t$}, {\em i.e.,} {\small $X^{ORA}(t)$} and {\small $Y^{ORA}(t)$}. The details of ORA are shown as Alg.1, which computes the fractional solution. Clearly, such fractional solutions constitute a basis of solution to the original problem, and later are used as inputs in next rounding algorithm.

\begin{algorithm}[!htp]
	\caption{An Online Regularization with look-Ahead Algorithm ({\bf ORA}) }
	\label{alg:framework}
	{\bf Input}: ${\small \eta}$, ${\small \varepsilon}$, ${\bm R}$, ${\bm C}$
	
	{\bf Output}: ${\small X^{ORA}(t)}$, ${\small X^{ORA}(t)}$
	
	{\bf Initialize}:  ${\small X^{ORA}(t)=0}$, ${\small Y^{ORA}(t)=0}$
	\begin{algorithmic}[1]
		\For{\small $t=-(L+1),\cdots,T$ }
		\State {\small $\pi\gets t$ mod $(L+1)$ and  $t^{(\pi)}\gets t$ }
		\If {{\small $t^{(\pi)}\leq 0$}}
		\State {\small Remove the third term in (\ref{eq:min3})}
		\EndIf
		\If {{\small $t^{(\pi)}+L \geq T$}}
		\State {\small Remove the fifth term in (\ref{eq:min3})}
		\EndIf
		\State {\small Invoke the interior point method to solve $P_{ORA}^{(\pi)}$ and obtain  $X^{(\pi)}(t^{(\pi)}:t^{(\pi)}+L)$ and  $Y^{(\pi)}(t^{(\pi)}:t^{(\pi)}+L)$}
		\While {\small $1 \leq t \leq T$} 
		\State {\small $X^{ORA}(t)=\frac{1}{L+1}\sum\limits_{\pi=0}^{L}X^{(\pi)}(t)$}
		\State {\small $Y^{ORA}(t)=\frac{1}{L+1}\sum\limits_{\pi=0}^{L}Y^{(\pi)}(t)$}
		\EndWhile
		\EndFor
	\end{algorithmic}
	\end{algorithm}

\subsection{Competitive Analysis }\label{sec:sub_Competitive_Analysis}

Next, we analyze the competitive performance of ORA according to an online primal-dual framework~\cite{primal-dual}. Driven by this, we introduce the dual problem of fractional problem, denoted by {\small$D$}, to bridge the offline minimization problem {\small$Cost$} and online minimization problem {\small$P_{ORA}$} by constructing the following inequalities:

\vspace{-6mm}
{\small
\begin{equation}
\begin{aligned}
    \label{eq:chain}
	C^{ORA}(1:T)  \overset{(a)}\leq  {r_1}  D^{ORA} (1:T) \overset{(b)}\leq {r_1} D^{opt}(1:T)  \overset{(c)}\leq {r_1} P^{opt}(1:T) \overset{(d)}\leq {r_1} Cost^{opt}(1:T),
\end{aligned}
\end{equation}}
\vspace{-6mm}

\noindent where {\small$Cost^{opt}$}, {\small$P^{opt}$} and {\small$D^{opt}$} denote the optimal objective function values of offline minimization problem {\small$Cost$}, fractional primal problem {\small$P$} and fractional dual problem {\small$D$}. Let {\small$D^{ORA}$} be an objective function value of fractional dual problem {\small$D$}, obtained by a constructed solution mapped from the fractional solution of online problem {\small$P_{ORA}$}. {\small$C^{ORA}$} is the objective function value of online minimization problem {\small$P_{ORA}$}.

As {\small $P$} is obtained from {\small $Cost$} via relaxing its integral constraints, inequality (\ref{eq:chain}d) holds. In addition, inequality (\ref{eq:chain}c) apparently follows from the Weak Duality Theorem~\cite{Convex}. Since the fractional dual problem {\small $D$} is a maximization problem and the library of online dual solutions obtained by ORA are feasible to problem {\small $D$}, we can consequently obtain inequality (\ref{eq:chain}b).

The rest of this subsection presents how to establish inequality (\ref{eq:chain}a). Specifically, we first formulate the fractional dual problem {\small $D$}. Then, we construct a dual solution for problem {\small $D^{ORA}$} and check the feasibility. Third, due to add two regularization terms in (\ref{eq:min3}), the cost gap between the online primal problem and online dual problem should be quantified. Finally, we obtain the competitive performance of ORA.

\vspace{3mm}
{\bf\em(1) Formulating Lagrange dual problem}.
The fractional dual problem {\small $D$} can be derived by the fractional minimization problem {\small $P$}, where {\small $\theta_{m,n,k}(t),\alpha_{n,k}(t),\beta_{m,k}(t),\rho_m(t),\mu_m(t)$} denote the Lagrangian dual variables corresponding to constraints (\ref{eq:min}a) - (\ref{eq:min}e)

\vspace{-3mm}
{\small
\begin{equation}
\label{eq:max}
\begin{aligned}
\text{maximize }	
\sum_{t}\sum_{n}\sum_{k}{\alpha}_{n,k}(t)-\sum_{t}\sum_{m}\rho_m(t)R_m-\sum_{t}\sum_{m}\mu_m(t)C_m
\end{aligned}
 \end{equation}	}
\vspace{-3mm}

{\small
subject to:  
\begin{align}
\sum_{n}\theta_{m,n,k}(t)-\beta_{m,k}(t)+{\beta}_{m,k}(t+1)-\rho_m(t)r_k\leq l_{m,k},\forall m,k,t, \tag{\ref{eq:max}a}\\
\alpha_{n,k}(t) \leq d_{m,n}\lambda_{n,k}(t)+ \theta_{m,n,k}(t)+\mu_m(t)\lambda_{n,k}(t)c_k, \forall m,n,k,t, \tag{\ref{eq:max}b }\\
 \beta_{m,k}(t)\leq b_{m,k}, \forall m,k,t,
 \tag{\ref{eq:max}c}\\
 \theta_{m,n,k}(t),\alpha_{n,k}(t),\beta_{j,k}(t),\rho_m(t),\mu_m(t)  \ge 0, \forall m,n,k,t, \tag{\ref{eq:max}d}
 \end{align}}
\vspace{-6mm}

\noindent where {\small$\forall m,n,k,t$} represent  {\small$\forall m\in\mathcal{M},n\in\mathcal{N}, k\in\mathcal{K}, t\in\mathcal{T}$}.

{\bf\em(2) Checking the fractional dual feasibility}.

Now, we concentrate on one version of ORA and explain the decision solutions obtained by all episodes of {\small $P_{ORA}^{(\pi)}$} are feasible to the fractional dual problem. Similar to (\ref{eq:max}), we can formulate the online dual problem of (\ref{eq:min3}) and let {\small $\theta_{m,n,k}^{(\pi)}(t),\alpha_{n,k}^{(\pi)}(t),\beta_{m,k}^{(\pi)}(t),\rho_m^{(\pi)}(t),\mu_m^{(\pi)}(t)$} denote the online dual variables corresponding to constraints (\ref{eq:min3}a) - (\ref{eq:min3}e). The objective function of (\ref{eq:min3}) has not contained the dynamic service placing overhead in the beginning time-frame {\small$t^{(\pi)}$}. To this end, we define the missing variables {\small$\beta_{m,k}^{(\pi)}(t^{(\pi)})$} as follows

\vspace{-3mm}
{\small
\begin{equation}
\label{eq:beta_1}
\begin{aligned}
  & \beta_{m,k}^{(\pi)}(t^{(\pi)})\triangleq \frac{b_{m,k}}{\eta}\ln\left(\frac{1+\frac{\varepsilon}{\mathcal{M}\mathcal{K}}}{x_{m,k}^{(\pi)}(t^{(\pi)}-1)+\frac{\varepsilon}{\mathcal{M}\mathcal{K}}}\right).
\end{aligned}
\end{equation}}
\vspace{-3mm}

{\bf\em Lemma 1}. The constructed solution {\small$\bm\theta^{(\pi)}(t)$}, {\small$\bm\alpha^{(\pi)}(t)$}, {\small$\bm\beta^{(\pi)}(t)$}, {\small$\bm\rho^{(\pi)}(t)$}, {\small$\bm\mu^{(\pi)}(t)$} from the online dual solution of {\small $P_{ORA}^{(\pi)}$} and (\ref{eq:beta_1}) are also feasible for fractional dual problem {\small$D$}.

Proof. See Appendix A.\qed
 
Lemma 1 can be proved by applying KKT optimality conditions~\cite{Convex} to {\small $P_{ORA}^{(\pi)}$}. In the following, we list the Complementary slackness and Optimality conditions in a disjunctive form, where {\small $c\perp d$} is equivalent to {\small$c,d \geq 0$} and  {\small$cd = 0$}.

\vspace{-3mm}
{\small
\begin{equation}
    \label{eq:Complementary}
      \text{\bf Complementary slackness}: 
\end{equation}}
\vspace{-3mm}
\vspace{-2mm}
{\small
\begin{align}
 \alpha_{n,k}^{(\pi)}(t)\perp\left [1-\sum_{m}  y_{m,n,k}^{(\pi)}(t)\right] = 0,\forall m,k,t, \tag{\ref{eq:Complementary}a}\\
\theta_{m,n,k}^{(\pi)}(t)\perp\left [x_{m,k}^{(\pi)}(t) - y_{m,n,k}^{(\pi)}(t)\right] = 0,\forall m,n,k,t, \tag{\ref{eq:Complementary}b}\\
\beta_{m,k}^{(\pi)}(t)\perp\left[ x_{m,k}^{(\pi)}(t) - x_{m,k}^{(\pi)}(t-1) - z_{m,k}^{(\pi)}(t)\right] = 0,\forall m,k,t, \tag{\ref{eq:Complementary}c}\\
\rho_m^{(\pi)}(t)\perp\left[\sum_{k}r_kx_{m,k}^{(\pi)}(t)-R_m\right] = 0,\forall m,t, \tag{\ref{eq:Complementary}d}\\
\mu_m^{(\pi)}(t)\perp\left[\sum_{k}\sum_{n}c_k\lambda_{n,k}(t) y_{m,n,k}^{(\pi)}(t)-C_m\right]=0,\forall m,t. \tag{\ref{eq:Complementary}e}
\end{align}}
\vspace{-3mm}

\vspace{-3mm}
{\small
\begin{equation}
    \label{eq:Stationarity}
      \text{\bf Stationarity/Optimality}:
\end{equation}}
\vspace{-4mm}

\vspace{-4mm}
{\small
\begin{align}
 x_{m,k}^{(\pi)}(t^{(\pi)})  \perp &\left[l_{m,k} +r_k \rho_m^{(\pi)}(t^{(\pi)})-\beta_{m,k}^{(\pi)}(t^{(\pi)}+1)\right.  \notag\\
 & \left.-\sum_{n} \theta_{m,n,k}^{(\pi)}(t^{(\pi)}) +\frac{b_{m,k}}{\eta}\ln\left(\frac{ 1 +\frac{\varepsilon}{\mathcal{M}\mathcal{K}}}{ x_{m,k}^{(\pi)}(t^{(\pi)}-1)+\frac{\varepsilon}{\mathcal{M}\mathcal{K}}}\right)\right] = 0,  \forall m,k, \tag{\ref{eq:Stationarity}a}
 \end{align}}
\vspace{-4mm}

\vspace{-4mm}
{\small
\begin{align}
 x_{m,k}^{(\pi)}(t) \perp  \left[ l_{m,k} +r_k \rho_m^{(\pi)}(t)-\beta_{m,k}^{(\pi)}(t+1)-\sum_{n} \theta_{m,n,k}^{(\pi)}(t) +\beta_{m,k}^{(\pi)}(t)\right] = 0,  \forall m,k,t \in [t^{(\pi)}+1,t^{(\pi)}+L-1], \tag{\ref{eq:Stationarity}b}
 \end{align}}
\vspace{-9mm}

\vspace{-9mm}
{\small
\begin{align}
 x_{m,k}^{(\pi)}(t^{(\pi)}+L)\perp & \left[ l_{m,k} +r_k \rho_m^{(\pi)}(t^{(\pi)}+L)+\beta_{m,k}^{(\pi)}(t^{(\pi)}+L)\right.  \notag\\
 & \left.-\sum_{n} \theta_{m,n,k}^{(\pi)}(t^{(\pi)}+L) -\frac{b_{m,k}}{\eta}\ln\left(\frac{ 1 +\frac{\varepsilon}{\mathcal{M}\mathcal{K}}}{ x_{m,k}^{(\pi)}(t^{(\pi)}+L)+\frac{\varepsilon}{\mathcal{M}\mathcal{K}}}\right)\right] = 0, \forall m,k, \tag{\ref{eq:Stationarity}c}
 \end{align}}
\vspace{-9mm}

\vspace{-9mm}
{\small
\begin{align}
 y_{m,n,k}^{(\pi)}(t)&(t) \perp \left[   \theta_{m,n,k}^{(\pi)}(t)+c_k\lambda_{n,k}(t)\mu_m^{(\pi)}(t)- \alpha_{n,k}^{(\pi)}(t) +d_{m,n} \lambda_{n,k}(t) \right] = 0, \forall m,n,k,t \in [t^{(\pi)},t^{(\pi)}+L], \tag{\ref{eq:Stationarity}d}
 \end{align}}
\vspace{-9mm}

\vspace{-9mm}
{\small
\begin{align}
  z_{m,k}^{(\pi)}(t) \perp \left[\beta_{m,k}^{(\pi)}(t)-b_{m,k}\right] = 0,\forall m,n,k,t \in [t^{(\pi)}+1,t^{(\pi)}+L].\tag{\ref{eq:Stationarity}e}
\end{align}}
\vspace{-3mm}

{\bf\em(3) Gauging the cost gap}.

Concentrating on a episode ({\em i.e.,} from {\small $t^{(\pi)}$} to {\small $t^{(\pi)}+L$}) of {\small$P_{ORA}^{(\pi)}$}, we introduce online primal overhead {\small $C^{(\pi)}(t^{(\pi)}:t^{(\pi)}+L)$} and online dual cost {\small $D^{(\pi)}(t^{(\pi)}:t^{(\pi)}+L)$} as follows

\vspace{-1mm}
{\small
\begin{equation*}
	C^{(\pi)}(t^{(\pi)}:t^{(\pi)}+L) \triangleq \sum\limits_{t=t^{(\pi)}}^{t^{(\pi)}+L}\bigg(C_R(t)+C_S(t)+C_D(t)\bigg),    
\end{equation*}}
\vspace{-3mm}

\vspace{-3mm}
{\small
\begin{equation*}
\begin{aligned}
	 D^{(\pi)}(t^{(\pi)}:t^{(\pi)}+L)  \triangleq \sum\limits_{t=t^{(\pi)}}^{t^{(\pi)}+L}\sum\limits_{n}\sum\limits_{k}{\alpha}_{j,k}^{(\pi)}(t)  -\sum\limits_{t=t^{(\pi)}}^{t^{(\pi)}+L}\sum\limits_{m}\rho_m^{(\pi)}(t)R_m-\sum\limits_{t=t^{(\pi)}}^{t^{(\pi)}+L}\sum\limits_{m}\mu_m^{(\pi)}(t)C_m.
\end{aligned}
\end{equation*}}
\vspace{-3mm}

\noindent The objective function in {\small $P_{ORA}^{(\pi)}$} has involved two regularized items, there thus be certain gap between online primal overhead and online dual overhead. In the following, we present the cost gap.

{\bf\em Lemma 2}. As for each {\small $ORA^{(\pi)}$}, we derive

\vspace{-2mm}
{\small
\begin{equation}
\label{eq:gap}
\begin{aligned}
	C^{(\pi)}(t^{(\pi)}:t^{(\pi)}+L) \leq  D^{(\pi)}(t^{(\pi)}:t^{(\pi)}+L) +\sum\limits_{m}\sum\limits_{k}{\Omega}_{m,k}^{(\pi)}(t^{(\pi)}) +\sum\limits_{m}\sum\limits_{k}{\phi}_{m,k}^{(\pi)}(t^{(\pi)})+\sum\limits_{m}\sum\limits_{k}{\psi}_{m,k}^{(\pi)}(t^{(\pi)}),
\end{aligned}
\end{equation}}	
\vspace{-2mm}
\noindent where the dual tail-terms {\small${\Omega}_{m,k}^{(\pi)}(t^{(\pi)})$}, {\small${\phi}_{m,k}^{(\pi)}(t^{(\pi)})$} and {\small${\psi}_{m,k}^{(\pi)}(t^{(\pi)})$} define as follows

\vspace{-2mm}
{\small
\begin{align}
	{\Omega}_{m,k}^{(\pi)}(t^{(\pi)}) \triangleq b_{m,k}{\left[x_{m,k}^{(\pi)}(t^{(\pi)})-x_{m,k}^{(\pi)}(t^{(\pi)}-1)\right]}^{+}, \tag{\ref{eq:gap}a}\\
	{\phi}_{m,k}^{(\pi)}(t^{(\pi)}) \triangleq -\frac{b_{m,k}}{\eta}x_{m,k}^{(\pi)}(t^{(\pi)})\ln\left(\frac{1+\frac{\varepsilon}{\mathcal{M}\mathcal{K}}}{x_{m,k}^{(\pi)}(t^{(\pi)}-1)+\frac{\varepsilon}{\mathcal{M}\mathcal{K}}}\right), \tag{\ref{eq:gap}b}\\
   {\psi}_{m,k}^{(\pi)}(t^{(\pi)}) \triangleq \frac{b_{m,k}}{\eta}x_{m,k}^{(\pi)}(t^{(\pi)}+L)\ln\left(\frac{1+\frac{\varepsilon}{\mathcal{M}\mathcal{K}}}{x_{m,k}^{(\pi)}(t^{(\pi)}+L)+\frac{\varepsilon}{\mathcal{M}\mathcal{K}}}\right). \tag{\ref{eq:gap}c}
\end{align}}
\vspace{-3mm}

Proof. See Appendix B. \qed

Note that the online problem {\small$P_{ORA}^{(\pi)}$} has not optimize the dynamic service placing cost in the first time-frame, which leads to dual tail-term {\small${\Omega}_{m,k}^{(\pi)}(t^{(\pi)})$}. Since the online primal objective in {\small$P_{ORA}^{(\pi)}$} includes two additional terms, the second and third tail-term are correspondingly obtained. Bounding cost gap is the main challenge of establishing inequality (\ref{eq:chain}a), we thus divide it into bi-steps.

{\bf\em Step 3-1} \emph {Bounding the dual tail-terms}.

Now, we are ready to bound three dual tail-terms, which originate from the same version {\small $\pi$} of {\small $ORA$}. The upper-bound of tail-terms are well-chosen portions of the online dual cost. In this context, we assume {\small$ \lceil r \rceil-1 < L$}, where {\small $ r \triangleq \max\limits_{m,k}(\frac{b_{m,k}}{l_{m,k}})$}. The proof is similar in the case of {\small$ \lceil r \rceil-1 \geq L$}.

{\bf\em Lemma 3}. For each {\small $ORA^{(\pi)}$}, the following two inequalities hold. 
\vspace{-2mm}
{\small
\begin{equation}
\label{eq:tail_1}
\begin{aligned}
	\sum\limits_{v=0}^{\lceil \frac{T}{L+1}\rceil}\sum\limits_{t^{(\pi)}=\pi\atop+(L+1)v}\sum\limits_{m}\sum\limits_{k}{\Omega}_{m,k}^{(\pi)}(t^{(\pi)}) \leq \eta(1+\varepsilon)\sum\limits_{v=0}^{\lceil \frac{T}{L+1}\rceil}\sum\limits_{t^{(\pi)}=\pi\atop+(L+1)v} D^{(\pi)}(t^{(\pi)}:t^{(\pi)}+\lceil r \rceil-1),
\end{aligned}
\end{equation}}	
\vspace{-2mm}
\vspace{-2mm}
{\small
\begin{equation}
\label{eq:tail_2}
\begin{aligned}
	\sum\limits_{v=-1}^{\lceil \frac{T}{L+1}\rceil}\sum\limits_{t^{(\pi)}=\pi\atop+(L+1)v}\sum\limits_{m}\sum\limits_{k}\left[{\phi}_{m,k}^{(\pi)}(t^{(\pi)}) + {\psi}_{m,k}^{(\pi)}(t^{(\pi)})\right] \leq \eta(1+\varepsilon)\sum\limits_{v=-1}^{\lceil \frac{T}{L+1}\rceil}\sum\limits_{t^{(\pi)}=\pi \atop +(L+1)v} D^{(\pi)}(t^{(\pi)}+L-\lceil r \rceil+1:t^{(\pi)}),
\end{aligned}
\end{equation}}	
\noindent where {\small$D^{(\pi)}(t)=0$} if {\small $t\leq 0$} or {\small $ t > T$}.

Proof. See Appendix C.\qed

Note that, according to (\ref{eq:tail_1}), the upper-bound of first dual tail-term {\small${\Omega}_{m,k}^{(\pi)}$} is a partial summation of online dual cost across intervals of size {\small$\lceil r \rceil$} at the beginning time-slot of each episode. The right-hand-side (RHS) of (\ref{eq:tail_2}) has a analogous explanation, while the partial summation is
replaced by intervals in the finish of every episode.

{\bf\em Step 3-2} \emph {Bounding the online dual overhead}.

The following lemma is to link the part of online dual overhead and the offline dual optimal overhead.

{\bf\em Lemma 4}. For each time interval {\small$[t_1,t_2]$}, we derive

{\small
\begin{equation}
\label{eq:interval}
\begin{aligned}
	 D^{(\pi)}(t_1:t_2) \leq & D^{opt}(t_1:t_2)-\sum\limits_{m}\sum\limits_{k}{\beta}_{m,k}^{opt}(t_1)x_{m,k}^{opt}(t_1-1)- \sum\limits_{m}\sum\limits_{k}{\beta}_{m,k}^{(\pi)}(t_2+1)x_{m,k}^{opt}(t_2) \\
	& +\sum\limits_{m}\sum\limits_{k}{\beta}_{m,k}^{opt}(t_2+1)x_{m,k}^{opt}(t_2)+\sum\limits_{m}\sum\limits_{k}{\beta}_{m,k}^{(\pi)}(t_1)x_{m,k}^{opt}(t_1-1),
\end{aligned}
\end{equation}}	
\noindent where {\small$x_{m,k}^{opt}(t)$}, {\small${\beta}_{m,k}^{opt}(t)$}, and {\small${\beta}_{m,k}^{(\pi)}(t)$} denote the optimal offline primal solution, the optimal offline dual solution, and the online dual solution.

{\bf Proof}:

For each time interval {\small $[t_1,t_2]$}, by applying the complementary slackness conditions to (\ref{eq:min}), we get
\vspace{-2mm}
{\small
\begin{equation}
\label{eq:interval_1}
\begin{aligned}
  Cost^{opt}(t_1:t_2)= &\sum\limits_{t=t_1}^{t_2}\sum\limits_{m}\sum\limits_{k}l_{m,k}x^{opt}_{m,k}(t) + \sum\limits_{t=t_1}^{t_2}\sum\limits_{m}\sum\limits_{n}\sum\limits_{k}d_{m,n}{\lambda_{n,k}(t)}{y^{opt}_{m,n,k}(t)}\\
  &+\sum\limits_{t=t_1}^{t_2}\sum\limits_{m}\sum\limits_{k}b_{m,k}z^{opt}_{m,k}(t) + \alpha_{n,k}^{opt}(t)\left[1-\sum_{m} y_{m,n,k}^{opt}(t)\right] + \theta_{m,n,k}^{opt}(t)\left[x_{m,k}^{opt}(t) - y_{m,n,k}^{opt}(t)\right]\\
  & + \beta_{m,k}^{opt}(t)\left[ x_{m,k}^{opt}(t) - x_{m,k}^{opt}(t-1) - z_{m,k}^{opt}(t)\right]  + \rho_m^{opt}(t)\left[\sum_{k}r_kx_{m,k}^{opt}(t)-R_m\right]\\
  & + \mu_m^{opt}(t)\left[\sum_{k}\sum_{n}c_k\lambda_{n,k}(t) y_{m,n,k}^{opt}(t)-C_m\right].
\end{aligned}
\end{equation}}	
\vspace{-6mm}

By rearranging terms in (\ref{eq:interval_1}), we have

\vspace{-2mm}
{\small
\begin{equation}
\label{eq:interval_2}
\begin{aligned}
   Cost^{opt}(t_1:t_2) 
   = & \sum\limits_{t=t_1}^{t_2}\sum\limits_{m}\sum\limits_{k}\alpha^{opt}_{n,k}(t)-\sum\limits_{t=t_1}^{t_2}\sum\limits_{m}\rho^{opt}_m(t)R_m-\sum\limits_{t=t_1}^{t_2}\sum\limits_{m}\mu^{opt}_m(t)C_m \\
  & - \sum\limits_{m}\sum\limits_{k}\beta_{m,k}^{opt}(t_1)x_{m,k}^{opt}(t_1-1)  + \sum\limits_{m}\sum\limits_{k}\beta_{m,k}^{opt}(t_2+1)x_{m,k}^{opt}(t_2) \\
  & +\sum\limits_{t=t_1}^{t_2}\sum\limits_{m}\sum\limits_{k}x_{m,k}^{opt}(t)\left[l_{m,k} -\sum_{n}\theta_{m,n,k}^{opt}(t)+r_k \rho_m^{opt}(t)+\beta_{m,k}^{opt}(t) -\beta_{m,k}^{opt}(t+1)\right] \\
  & + \sum\limits_{t=t_1}^{t_2}\sum\limits_{m}\sum\limits_{n}\sum\limits_{k} y_{m,n,k}^{opt}(t)\cdot \left[ \theta_{m,n,k}^{opt}(t)+c_k\lambda_{n,k}(t)\mu_m^{opt}(t)- \alpha_{n,k}^{opt}(t) +d_{m,n}\lambda_{n,k}(t)\right]\\
  & + \sum\limits_{t=t_1}^{t_2}\sum\limits_{m}\sum\limits_{k} z_{m,k}^{opt}(t) \left[\beta_{m,k}^{opt}(t)-b_{m,k}\right].
\end{aligned}
\end{equation}}	
\vspace{-6mm}

\noindent Utilizing the optimality conditions to (\ref{eq:interval_2}), we get

\vspace{-3mm}
{\small
\begin{equation}
\label{eq:interval_3}
\begin{aligned}
   Cost^{opt}(t_1:t_2)  = & \sum\limits_{t=t_1}^{t_2}\sum\limits_{m}\sum\limits_{k}\alpha^{opt}_{n,k}(t)-\sum\limits_{t=t_1}^{t_2}\sum\limits_{m}\rho^{opt}_m(t)R_m+\sum\limits_{m}\sum\limits_{k}{\beta}_{m,k}^{opt}(t_2+1)x_{m,k}^{opt}(t_2)\\
  & -\sum\limits_{t=t_1}^{t_2}\sum\limits_{m}\mu^{opt}_m(t)C_m -\sum\limits_{m}\sum\limits_{k}{\beta}_{m,k}^{opt}(t_1)x_{m,k}^{opt}(t_1-1).
\end{aligned}
\end{equation}}
\vspace{-1mm}

\noindent Based on the primal constraints (\ref{eq:min}b) and (\ref{eq:min}c), we have

\vspace{-1mm}
{\small
\begin{equation}
\label{eq:interval_4}
\begin{aligned}
   Cost^{opt}(t_1:t_2)  \geq &  \sum\limits_{t=t_1}^{t_2}\sum\limits_{m}\sum\limits_{k}l_{m,k}x_{m,k}^{opt}(t)
   +\sum\limits_{t=t_1}^{t_2}\sum\limits_{m}\sum\limits_{n}\sum\limits_{k}d_{m,n}{\lambda_{n,k}(t)}{y_{m,n,k}^{opt}}(t) +\sum\limits_{t=t_1}^{t_2}\sum\limits_{m}\sum\limits_{k}b_{m,k}z_{m,k}^{opt}(t)\\
  & + \sum\limits_{t=t_1}^{t_2}\sum\limits_{n}\sum\limits_{k}\alpha_{n,k}^{(\pi)}(t)\left[1-\sum_{m}  y_{m,n,k}^{opt}(t)\right]
   + \sum\limits_{t=t_1}^{t_2}\sum\limits_{m}\sum\limits_{n}\sum\limits_{k}\theta_{m,n,k}^{(\pi)}(t)\left[x_{m,k}^{opt}(t) -y_{m,n,k}^{opt}(t)\right]\\
  & + \sum\limits_{t=t_1}^{t_2}\sum\limits_{m}\sum\limits_{k}\beta_{m,k}^{(\pi)}(t)\left[ x_{m,k}^{opt}(t) - x_{m,k}^{opt}(t-1) - z_{m,k}^{opt}(t)\right]
   + \sum\limits_{t=t_1}^{t_2}\sum\limits_{m}\rho_m^{(\pi)}(t)\left[\sum_{k}r_kx_{m,k}^{opt}(t) -R_m\right]\\
  & + \sum\limits_{t=t_1}^{t_2}\sum\limits_{m}\mu_m^{(\pi)}(t)\left[\sum_{n}\sum_{k}c_k\lambda_{n,k}(t) y_{m,n,k}^{(\pi)}(t)- C_m\right].
\end{aligned}
\end{equation}}
\vspace{-1mm}

By rearranging the terms in (\ref{eq:interval_4}), we have

\vspace{-1mm}
{\small
\begin{equation*}
\label{eq:interval_5}
\begin{aligned}
   C^{opt}(t_1:t_2) \geq &  \sum\limits_{t=t_1}^{t_2}\sum\limits_{m}\sum\limits_{k}\alpha^{(\pi)}_{n,k}(t)-\sum\limits_{t=t_1}^{t_2}\sum\limits_{m}\rho^{(\pi)}_m(t)R_m-\sum\limits_{t=t_1}^{t_2}\sum\limits_{m}\mu^{(\pi)}_m(t)C_m -\sum\limits_{m}\sum\limits_{k}{\beta}_{m,k}^{(\pi)}(t_1)x_{m,k}^{opt}(t_1-1)\\
  & +\sum\limits_{m}\sum\limits_{k}{\beta}_{m,k}^{(\pi)}(t_2+1)x_{m,k}^{opt}(t_2)+ \sum\limits_{t=t_1}^{t_2}\sum\limits_{m}\sum\limits_{k}z_{m,k}^{opt}(t)\left[\beta_{m,k}^{(\pi)}(t)-b_{m,k}\right]\\
  & + \sum\limits_{t=t_1}^{t_2}\sum\limits_{m}\sum\limits_{k}x_{m,k}^{opt}(t) \cdot \left[l_{m,k} -\sum_{n} \theta_{m,n,k}^{(\pi)}(t)  + r_k \rho_m^{(\pi)}(t)+\beta_{m,k}^{(\pi)}(t)-\beta_{m,k}^{(\pi)}(t+1)\right]\\
  & + \sum\limits_{t=t_1}^{t_2}\sum\limits_{m}\sum\limits_{n}\sum\limits_{k} y_{m,n,k}^{opt}(t) \cdot \left[ \theta_{m,n,k}^{(\pi)}(t)+c_k\lambda_{n,k}(t)\mu_m^{(\pi)}(t)- \alpha_{n,k}^{(\pi)}(t)+d_{m,n}\lambda_{n,k}(t)\right].
\end{aligned}
\end{equation*}}
\vspace{-1mm}

\noindent Note that {\small $x_{m,k}^{opt}(t)$} and {\small $y_{m,n,k}^{opt}(t)$} are non-negative. And, the optimality conditions indicate that, to prevent the objective value of online dual problem from going down to minus infinity, constrains (\ref{eq:max}a), (\ref{eq:max}b) and (\ref{eq:max}c) should be respected. Thus, we get

\vspace{-1mm}
{\small
\begin{equation}
\label{eq:interval_6}
\begin{aligned}
  Cost^{opt}(t_1:t_2)  & \geq \sum\limits_{t=t_1}^{t_2}\sum\limits_{m}\sum\limits_{k}\alpha^{(\pi)}_{n,k}(t)-\sum\limits_{t=t_1}^{t_2}\sum\limits_{m}\rho^{(\pi)}_m(t)R_m-\sum\limits_{t=t_1}^{t_2}\sum\limits_{m}\mu^{(\pi)}_m(t)C_m \\
  & -\sum\limits_{m}\sum\limits_{k}{\beta}_{m,k}^{(\pi)}(t_1)x_{m,k}^{opt}(t_1-1)+\sum\limits_{m}\sum\limits_{k}{\beta}_{m,k}^{(\pi)}(t_2+1)x_{m,k}^{opt}(t_2).
\end{aligned}
\end{equation}}
\vspace{-1mm}

\noindent Based on (\ref{eq:interval_3}) and (\ref{eq:interval_6}), we get 

\vspace{-1mm}
{\small
\begin{equation*}
\label{eq:interval_5}
\begin{aligned}
  & \sum\limits_{t=t_1}^{t_2}\sum\limits_{m}\sum\limits_{k}\alpha^{(\pi)}_{n,k}(t)-\sum\limits_{t=t_1}^{t_2}\sum\limits_{m}\rho^{(\pi)}_m(t)R_m-\sum\limits_{t=t_1}^{t_2}\sum\limits_{m}\mu^{(\pi)}_m(t)C_m \\
  & -\sum\limits_{m}\sum\limits_{k}{\beta}_{m,k}^{(\pi)}(t_1)x_{m,k}^{opt}(t_1-1)+\sum\limits_{m}\sum\limits_{k}{\beta}_{m,k}^{(\pi)}(t_2+1)x_{m,k}^{opt}(t_2)\\
  & \leq \sum\limits_{t=t_1}^{t_2}\sum\limits_{m}\sum\limits_{k}\alpha^{opt}_{n,k}(t)-\sum\limits_{t=t_1}^{t_2}\sum\limits_{m}\rho^{opt}_m(t)R_m-\sum\limits_{t=t_1}^{t_2}\sum\limits_{m}\mu^{opt}_m(t)C_m \\
  & -\sum\limits_{m}\sum\limits_{k}{\beta}_{m,k}^{(\pi)}(t_1)x_{m,k}^{opt}(t_1-1)+\sum\limits_{m}\sum\limits_{k}{\beta}_{m,k}^{(\pi)}(t_2+1)x_{m,k}^{opt}(t_2).
\end{aligned}
\end{equation*}}
\vspace{-1mm}
\qed

{\bf\em(4) Obtaining the competitive ratio of ORA}.

Next, we analyze the competitive performance of ORA.

{\bf Theorem 1}. With {\small$L \geq 1$} and {\small$\lceil r\rceil < L+1$}, the competitive ratio of ORA can be computed as follows

\vspace{-6mm}
{\small
\begin{equation*}
\label{eq:r_1}
r_1=
1+\frac{3\eta(1+\frac{\varepsilon}{\mathcal{M}\mathcal{K}})\lceil r\rceil}{L+1}.
\end{equation*}}
\vspace{-6mm}

Before proving Theorem 1, we calculate the total cost of ORA as follows,

\vspace{-3mm}
{\small
\begin{equation}
\label{eq:ORA}
\begin{aligned}
  C^{ORA}(1:T)= \sum\limits_{t=1}^{T}\sum\limits_{m}\sum\limits_{k}l_{m,k}x^{ORA}_{m,k}(t)& + \sum\limits_{t=1}^{T}\sum\limits_{m}\sum\limits_{n}\sum\limits_{k}d_{m,n}{\lambda_{n,k}(t)}{y^{ORA}_{m,n,k}(t)}\\
  &+\sum\limits_{t=1}^{T}\sum\limits_{m}\sum\limits_{k}b_{m,k}{\left[x_{m,k}^{ORA}(t)-x_{m,k}^{ORA}(t-1)\right]}^{+},
\end{aligned}
\end{equation}}	
\vspace{-3mm}

\noindent where {\small $x^{ORA}_{m,k}(t)$} and {\small $y^{ORA}_{m,n,k}(t)$} can be obtained by Alg.1. By utilizing Jensen's Inequality, we can derive

\vspace{-6mm}
{\small
\begin{equation}
\label{eq:ORA_1}
\begin{aligned}
  C^{ORA}(1:T)\leq \frac{1}{L+1} \sum\limits_{\pi=0}^{L}C^{(\pi)}(1:T).
\end{aligned}
\end{equation}}	
\vspace{-3mm}

\noindent By applying Lemma 2 to (\ref{eq:ORA_1}), the upper-bound of ORA {\small$C^{ORA}(1:T)$} is

\vspace{-3mm}
{\small
\begin{equation}
\label{eq:ORA_2}
\begin{aligned}
   C^{ORA}(1:T)   
   \leq & \frac{1}{L+1} \sum\limits_{\pi=0}^{N}\sum\limits_{v=-1}^{\lceil \frac{T}{L+1}\rceil}\sum\limits_{t^{(\pi)}=\pi\atop+(L+1)v}\bigg\{D^{(\pi)}(t^{(\pi)}:t^{(\pi)}+L)\bigg.\\
  & +\sum\limits_{m}\sum\limits_{k}{\Omega}_{m,k}^{(\pi)}(t^{(\pi)})+\sum\limits_{m}\sum\limits_{k}{\phi}_{m,k}^{(\pi)}(t^{(\pi)}) +\sum\limits_{m}\sum\limits_{k}{\psi}_{m,k}^{(\pi)}(t^{(\pi)})\bigg\}.
\end{aligned}
\end{equation}}	
\vspace{-3mm}

\noindent Based on Lemma 1, the online dual cost in (\ref{eq:ORA_2}) amounts to 

\vspace{-3mm}
{\small
\begin{equation}
\label{eq:ORA_3}
\begin{aligned}
   & \frac{1}{L+1}\sum\limits_{\pi=0}^{L}D^{(\pi)}(1:T) \leq D^{opt}(1:T).
\end{aligned}
\end{equation}}	
\vspace{-3mm}

\noindent The RHS of (\ref{eq:ORA_2}) only remains three dual tail-terms that need to bound.

Recall that Lemma 3, we can deduce two important conclusions, and list as follows (the proof is given by Appendix D).

(i) The upper-bound of {\small$\sum\limits_{m}\sum\limits_{k}{\Omega}_{m,k}^{(\pi)}(t^{(\pi)})$} is 

\vspace{-3mm}
{\small
\begin{equation}
\label{eq:ORA_4}
\left\{ 
\begin{aligned}
\eta(1+\frac{\varepsilon}{\mathcal{M}\mathcal{K}}) \sum\limits_{t=t^{(\pi)}}^{t^{(\pi)}-1\atop+\lceil r\rceil} D^{(\pi)}(t) & , &  t^{(\pi)} \in [1,T-\lceil r\rceil], \\
\eta(1+\frac{\varepsilon}{\mathcal{M}\mathcal{K}}) \sum\limits_{t=t^{(\pi)}}^{T}D^{(\pi)}(t) & , &   t^{(\pi)} \in [T-\lceil r\rceil+1,T],
\end{aligned}
\right.
\end{equation}}
\vspace{-6mm}

\noindent where {\small $D^{(\pi)}(t)=\sum\limits_{n}\sum\limits_{k}{\alpha}_{n,k}^{(\pi)}(t)-\sum\limits_{m}\rho_m^{(\pi)}(t)R_m-\sum\limits_{m}\mu_m^{(\pi)}(t)C_m$}. For simplicity, we use {\small$D^{(\pi)}(t)$} in the following proof.

(ii) The upper-bound of {\small$\sum\limits_{m}\sum\limits_{k}\left[{\phi}_{m,k}^{(\pi)}(t^{(\pi)})+{\psi}_{m,k}^{(\pi)}(t^{(\pi)})\right]$} is 

\vspace{-3mm}
{\small
\begin{equation}
\label{eq:ORA_5}
\left\{ 
\begin{aligned}
\eta(1+\frac{\varepsilon}{\mathcal{M}\mathcal{K}})\sum\limits_{t=t^{(\pi)}+L\atop -\lceil r\rceil+1}^{t^{(\pi)}+L}D^{(\pi)}(t)& , &  t^{(\pi)} \in [-L+\lceil r\rceil+1,T-L], \\
\eta(1+\frac{\varepsilon}{\mathcal{M}\mathcal{K}})\sum\limits_{t=1}^{t^{(\pi)}+L}D^{(\pi)}(t) & , &  t^{(\pi)} \in [-L+1,-L+\lceil r\rceil].
\end{aligned}
\right.
\end{equation}}
\vspace{-2mm}

{\bf\em Proof of Theorem 1.}

Based on Lemma 4 and applying (\ref{eq:ORA_4})-(\ref{eq:ORA_5}) to (\ref{eq:tail_1})-(\ref{eq:tail_2}) in Lemma 3, we obtain

\vspace{-3mm}
{\small
\begin{equation}
\label{eq:ORA_10}
\begin{aligned}
   C^{ORA}(1:T)
  \leq  & \frac{1}{L+1} \sum\limits_{\pi=0}^{L}D^{(\pi)}(1:T)\\
  & +\frac{1}{L+1}\eta(1+\frac{\varepsilon}{\mathcal{M}\mathcal{K}})\cdot \bigg\{ \sum\limits_{t^{(\pi)}=1}^{T-\lceil r\rceil}D^{(\pi)}(t^{(\pi)}:t^{(\pi)}+\lceil r\rceil -1)  
  + \sum\limits_{t^{(\pi)}=T\atop-\lceil r\rceil+1}^{T}D^{(\pi)}(t^{(\pi)}:T)\\
  & +\sum\limits_{t^{(\pi)}=-(L+1)}^{-L+\lceil r\rceil}D^{(\pi)}(1:t^{(\pi)}+L) +\sum\limits_{t^{(\pi)}=-L\atop+\lceil r\rceil+1}^{T-L}D^{(\pi)}(t^{(\pi)}+L-\lceil r\rceil+1:t^{(\pi)}+L) \bigg\} .
\end{aligned}
\end{equation}}	
\vspace{-3mm}

\noindent For each time-slot {\small $t$}, inequality {\small $D^{(\pi)}(t)\geq 0$} holds. Therefore, from (\ref{eq:ORA_10}), we get

\vspace{-2mm}
{\small
\begin{equation}
\label{eq:ORA_11}
\begin{aligned}
  C^{ORA}(1:T)
  \leq  & \frac{1}{L+1} \sum\limits_{\pi=0}^{L}D^{(\pi)}(1:T)+\frac{1}{L+1}\eta(1+\frac{\varepsilon}{\mathcal{M}\mathcal{K}})\cdot \bigg\{\sum\limits_{t^{(\pi)}=1}^{T-\lceil r\rceil}D^{(\pi)}(1:t^{(\pi)}) \bigg.\\
  & + \sum\limits_{t^{(\pi)}=1}^{T-\lceil r\rceil}D^{(\pi)}(t^{(\pi)}:t^{(\pi)}+\lceil r\rceil -1) +\sum\limits_{t^{(\pi)}=T\atop-\lceil r\rceil+1}^{T}D^{(\pi)}(t^{(\pi)}:T)+\sum\limits_{t^{(\pi)}=-L+1}^{-L+\lceil r\rceil}D^{(\pi)}(1:t^{(\pi)}+L)\\
  & +\sum\limits_{t^{(\pi)}=-L\atop +\lceil r\rceil+1}^{T-L}D^{(\pi)}(t^{(\pi)}+L-\lceil r\rceil+1:t^{(\pi)}+L) 
  + \sum\limits_{t^{(\pi)}=-L+\atop\lceil r\rceil+2}^{T}D^{(\pi)}(t^{(\pi)}:T)\bigg\} .
\end{aligned}
\end{equation}}	
\vspace{-2mm}

\noindent According to Lemma 4, we derive 

\vspace{-6mm}
{\small
\begin{equation}
\label{eq:ORA_12}
\begin{aligned}
   C^{ORA}(1:T)
  \leq  & \frac{1}{L+1} \sum\limits_{\pi=0}^{L}D^{(\pi)}(1:T)+\frac{1}{L+1}\eta(1+\frac{\varepsilon}{\mathcal{M}\mathcal{K}})\cdot \bigg\{ 2\lceil r\rceil D^{opt}(1:T) \\
  & + \sum\limits_{t^{(\pi)}=1}^{T-\lceil r\rceil}\sum\limits_{m}\sum\limits_{k}{\beta}_{m,k}^{(\pi)}(t^{(\pi)})x_{m,k}^{opt}(t^{(\pi)}-1) + \sum\limits_{t^{(\pi)}=T\atop-\lceil r\rceil+1}^{T}\sum\limits_{m}\sum\limits_{k}{\beta}_{m,k}^{(\pi)}(t^{(\pi)})x_{m,k}^{opt}(t^{(\pi)}-1)\\
  & + \left.\sum\limits_{t^{(\pi)}=-L\atop +\lceil r\rceil+1}^{T-L}\sum\limits_{m}\sum\limits_{k} {\beta}_{m,k}^{(\pi)}(t^{(\pi)}+L-\lceil r\rceil+1)\cdot x_{m,k}^{opt}(t^{(\pi)}+L-\lceil r\rceil) + \sum\limits_{t^{(\pi)}=-L\atop+\lceil r\rceil+2}^{T}\sum\limits_{m}\sum\limits_{k} {\beta}_{m,k}^{(\pi)}(t^{(\pi)})x_{m,k}^{opt}(t^{(\pi)}-1)\right.\\
  & \left. - \sum\limits_{t^{(\pi)}=1}^{\lceil r\rceil-1}\sum\limits_{m}\sum\limits_{k}{\beta}_{m,k}^{(\pi)}(t^{(\pi)}+1)x_{m,k}^{opt}(t^{(\pi)}) - \sum\limits_{t^{(\pi)}=1}^{T-\lceil r\rceil}\sum\limits_{m}\sum\limits_{k}{\beta}_{m,k}^{(\pi)}(t^{(\pi)}+\lceil r\rceil)x_{m,k}^{opt}(t^{(\pi)}+\lceil r\rceil-1)\right.\\ 
  & - \sum\limits_{t^{(\pi)}=-L+1}^{-L+\lceil r\rceil}\sum\limits_{m}\sum\limits_{k} {\beta}_{m,k}^{(\pi)}(t^{(\pi)}+L+1)x_{m,k}^{opt}(t^{(\pi)}+L)\\
  & - \sum\limits_{t^{(\pi)}=-L\atop+\lceil r\rceil+1}^{T-L}\sum\limits_{m}\sum\limits_{k} {\beta}_{m,k}^{(\pi)}(t^{(\pi)}+L+1)x_{m,k}^{opt}(t^{(\pi)}+L)\bigg\}. 
\end{aligned}
\end{equation}}	
\vspace{-6mm}

As constraint (\ref{eq:max}c) states {\small$ 0 \leq \beta_{m,k}^{(\pi)}(t)\leq b_{m,k}$}, we have 
\vspace{-2mm}
{\small
\begin{equation}
\label{eq:ORA_13}
\begin{aligned}
   C^{ORA}(1:T)
  \leq & \frac{1}{L+1} \sum\limits_{\pi=0}^{L}D^{(\pi)}(1:T)+\frac{1}{L+1}\eta(1+\frac{\varepsilon}{\mathcal{M}\mathcal{K}})\cdot \bigg\{ 2\lceil r\rceil D^{opt}(1:T) \\
  & + \sum\limits_{t^{(\pi)}=1}^{T}\sum\limits_{m}\sum\limits_{k}{\beta}_{m,k}^{(\pi)}(t^{(\pi)})x_{m,k}^{opt}(t^{(\pi)}-1) \\
  & \left.+ \sum\limits_{t^{(\pi)}=1}^{T-\lceil r\rceil}\sum\limits_{m}\sum\limits_{k}b_{m,k}x_{m,k}^{opt}(t^{(\pi)}) + \sum\limits_{t^{(\pi)}=T\atop-\lceil r\rceil+1}^{T-1}\sum\limits_{m}\sum\limits_{k}b_{m,k}x_{m,k}^{opt}(t^{(\pi)}) \right.\\
  & - \sum\limits_{t^{(\pi)}=2}^{\lceil r\rceil+1}\sum\limits_{m}\sum\limits_{k}{\beta}_{m,k}^{(\pi)}(t^{(\pi)})x_{m,k}^{opt}(t^{(\pi)}-1)  - \sum\limits_{t^{(\pi)}=\lceil r\rceil+ 2}^{T}\sum\limits_{m}\sum\limits_{k}{\beta}_{m,k}^{(\pi)}(t^{(\pi)})x_{m,k}^{opt}(t^{(\pi)}-1) \bigg\}. 
\end{aligned}
\end{equation}}	
\vspace{-6mm}

As {\small $x_{m,k}^{(\pi)}(0)=0 $} and {\small $x_{m,k}^{opt}(t)\geq 0 $}, from (\ref{eq:ORA_13}), we obtain  

\vspace{-3mm}
{\small
\begin{equation}
\label{eq:ORA_14}
\begin{aligned}
   C^{ORA}(1:T)
  \leq   \frac{1}{L+1} \sum\limits_{\pi=0}^{L}D^{(\pi)}(1:T)+\frac{1}{L+1}\eta(1+\frac{\varepsilon}{\mathcal{M}\mathcal{K}})
  \cdot \left[ 2\lceil r\rceil D^{opt}(1:T) + \sum\limits_{t=1}^{T}\sum\limits_{m}\sum\limits_{k}l_{m,k}x_{m,k}^{opt}(t)\right]. 
\end{aligned}
\end{equation}}	

\noindent Note that {\small$b_{m,k}x_{m,k}^{opt}(t) \leq \lceil r\rceil b_{m,k}x_{m,k}^{opt}(t)$}. Lastly, based on Lemma 1 and the duality theorem~\cite{Convex}, from (\ref{eq:ORA_14}), we derive

\vspace{-3mm}
{\small
\begin{equation*}
\label{eq:ORA_15}
\begin{aligned}
   C^{ORA}(1:T)
   & \leq    C^{opt}(1:T) + \frac{1}{L+1}\eta(1+\frac{\varepsilon}{\mathcal{M}\mathcal{K}})
  \cdot \{ 2\lceil r\rceil C^{opt}(1:T)+\lceil r\rceil C^{opt}(1:T)\} \\
  & \leq    \bigg \{ 1+ \frac{3\eta(1+\frac{\varepsilon}{\mathcal{M}\mathcal{K}})\lceil r\rceil}{L+1}\bigg\} C^{opt}(1:T) .
\end{aligned}
\end{equation*}}
\vspace{-3mm}
\qed

\section{Online Dependent Rounding Algorithm }\label{sec:rounding_algorithm}
The online regularization with look-ahead algorithm calculates fractional solutions for fractional minimization problem {\small $P$} in each time slot. When considering the integral constraints of optimization variables, we should translate fractional solution {\small $\bm {\widetilde{x}}$ } into integer ones {\small $\bm{\overline{x}}$}, which simultaneously satisfies feasibility constraints (\ref{eq:min}a), (\ref{eq:min}d) and (\ref{eq:min}g).

\subsection{Algorithm design}\label{sec:Algorithm_Design}

A natural approach is the independent rounding method~\cite{independent} whose key idea is that each control solution is rounded up or down individually. Yet, such simple solution may obtain a infeasible solution or lead to a high cost and even system instability. To this end, we design a randomized dependent service placing algorithm {\small $RDSP$}, enlightened by the dependent rounding technique~\cite{Rounding}. The online algorithm {\small $RDSP$} employs the interdependence of {\small $\bm {\widetilde{x}}$ }, consisting of the following step:

{\bf\em Step 1.} We first construct a bipartite graph;

{\bf\em Step 2.} We apply the dependent rounding method to calculate the rounded solution.

\noindent This algorithm executes a series of rounding iterations. Let {\small $x_{m,k}^{h}(t)$} denote solution {\small $x_{m,k}(t)$} after {\small$h$}-th rounding iteration. 

We begin with the {\small$h+1$} rounding iteration. If a service is fractionally allocated to more than one SBS, we refer it as a floating service. Similarly, a SBS can be called as a floating SBS, if it has more than one floating service dispatched to it in the current rounding iteration. Let {\small $K_f$} and {\small$M_f$} denote the set of floating services and floating SBSs in the current iteration, respectively. 
We construct a simple bipartite graph {\small$G=(M_f,K_f,E)$}, whose vertexs are floating SBSs and services, and edges correspond to current fractional solutions, {\em i.e.,} {\small $E=\{ (m,k): x_{m,k}(t) \in (0,1) \}$}. We adopt the Depth-First-Search method to find a circle or longest path that partitions the edges into bi-matching sets {\small $L_1$} and {\small $L_2$}.
The details of randomized dependent service placing algorithm {\small $RDSP$}  is shown in Alg.2, which computes integer solutions.

\vspace{-2mm}
\begin{algorithm}[!htp]
	\caption{The Randomized Dependent Service Placing Algorithm ({\bf RDSP}) }
	\label{alg:framework}
	{\bf Input}: $ \bm{\widetilde {x}}(t)$
	
	{\bf Output}:  $\bm{\overline{x}}(t)$

	\begin{algorithmic}[1]
	\State {\small Construct bipartite graph $G=(M_f,K_f,E)$};
	\While {$E$ is not empty} 
	\While {there exists a cycle or longest path}
	\State Partition cycle/path into two matchings {\small $L_1$} and {\small $L_2$};
		\State Let {\small $\xi= \min \{ \alpha > 0 \mid \exists(m,k)\in L_1 : (x_{m,k}(t)+\alpha=1) \vee \exists(m,k)\in L_2: (x_{m,k}(t)-\alpha=0)  \}$};
		\State Let {\small $\omega= \min \{ \alpha > 0 \mid \exists(m,k)\in L_1 : (x_{m,k}(t)-\alpha=0) \vee \exists(m,k)\in L_2: (x_{m,k}(t)+\alpha=1)  \}$};
		\State Set {\small
		$x_{m,k}(t)=x_{m,k}(t)+{\xi}, \forall (m,k) \in L_1 $} and {\small
		$x_{m,k}(t)=x_{m,k}(t)-{\xi}, \forall (m,k) \in L_2 $} with probability {\small ${\omega}/({\omega}+{\xi})$};
		\State Set {\small
		$x_{m,k}(t)=x_{m,k}(t)-{\omega}, \forall (m,k) \in L_1 $} and {\small
		$x_{m,k}(t)=x_{m,k}(t)+{\omega}, \forall (m,k) \in L_2 $} with probability {\small ${\xi}/({\omega}+{\xi})$};
		\State Remove edge {\small$(m,k)$} from {\small$E$} if {\small $x_{m,k}(t)$} is integer; 
	  \EndWhile
	  \EndWhile
	\end{algorithmic}
	\end{algorithm}
\vspace{4mm}

\subsection{Algorithm Analysis }\label{sec:Algorithm_Analysis}

Now, we prove the inequality
{\small $C^{ORA}(1:T) \geq \frac {1}{r_2}C^{RDSP}(1:T)$}, where {\small$C^{ORA}(1:T)$} and {\small$C^{RDSP}(1:T)$} represent the objective value of online primal problem and online rounding problem, achieved by the fractional solution by {\small$ORA$} and the rounded solution by {\small$RDSP$}.

{\bf\em Lemma 5}. {\small$RDSP$} guarantees {\small $\mathbb{E}[\overline{x}_{m,k}(t)]=\widetilde {x}_{m,k}(t), \forall i,k,t$}.

Proof:
Focusing on {\small$h+1$} rounding iteration, we obtain
\vspace{-3mm}
{\small
\begin{align*}
	\mathbb{E}[{x}_{m,k}^{h+1}(t)]=\dfrac{\omega}{\omega+\xi}({x}_{m,k}^{h}(t)+\xi)+\dfrac{\xi}{\omega+\xi}({x}_{m,k}^{h}(t)-\omega),
\end{align*}}
\vspace{-3mm}
or
\vspace{-3mm}
{\small
\begin{align*}
	\mathbb{E}[{x}_{m,k}^{h+1}(t)]=\dfrac{\omega}{\omega+\xi}({x}_{m,k}^{h}(t)-\xi)+\dfrac{\xi}{\omega+\xi}({x}_{m,k}^{h}(t)+\omega).
\end{align*}}
\qed
\vspace{-2mm}

{\bf Theorem 2}. The proposed {\small$RDSP$} algorithm can achieve  competitive ratio {\small$r_2$} compared with {\small$C^{ORA}(1:T) $}.

Proof: 

\vspace{-6mm}
{\small
\begin{align}
\label{eq:chain_1}
    \mathbb{E}\left[C^{RDSP}(1:T)\right] & =\mathbb{E}\left[\sum_{t}\sum_{m}\sum_{k}l_{m,k}\overline{x}_{m,k}(t)+ \sum_{t}\sum_{m}\sum_{n}\sum_{k}d_{m,n}{\lambda_{n,k}(t)}{y_{m,n,k}(t)}+ \sum_{t}\sum_{m}\sum_{k}b_{m,k}{z_{m,k}(t)} \right]\notag\\
	& \leq C^{ORA}(1:T)+C^{ORA}(1:T) +\sum_{t}\sum_{m}\sum_{k}\max\limits_{m,k}(\frac{b_{m,k}}{l_{m,k}})l_{m,k}\widetilde{x}_{m,k}(t)\notag\\
	& \leq (2+\max \lceil r \rceil)C^{ORA}(1:T).
\end{align}}
\vspace{-2mm}
\qed

Together (\ref{eq:chain}) with (\ref{eq:chain_1}), we achieve overall competitive ratio:

\vspace{-3mm}
{\small
\begin{equation}
\begin{aligned}
    \label{eq:chain_2}
	 & C^{RDSP}(1:T) \leq {r_2}C^{ORA}(1:T) \leq {r_1}{r_2} Cost^{opt}(1:T),
\end{aligned}
\end{equation}}
\vspace{-3mm}

\noindent where {\small$r_1=
1+\frac{3\eta(1+\frac{\varepsilon}{\mathcal{M}\mathcal{K}})\lceil r\rceil}{L+1}$} and {\small $r_2=2+\max \lceil r \rceil$}.

{\bf Theorem 3.} $ORA$ is a polynomial running time algorithm with computational complexity {\small $O(TK^2M^2)$}.

{\em Proof.} See Appendix E.\qed
 \section{Numerical Experiment }\label{sec:simulation}
 
In this section, we further conduct the simulation to evaluate the effectiveness of our proposed algorithm. Our goals are threefold:

(1) To evaluate the performance of our proposed algorithm under realistic requests trace;

(2) To understand which factors influence the performance of our proposed algorithm;

(3) To verify whether our online algorithm is compatible with different system settings.

\subsection{Experimental setup}
Now, we discuss the details of simulation scenario.

1) MEC network

Similarly to the previous related work~\cite{KP}, we also adopt the similar network structure. {\small$M=5$} SBSs are orderly deployed on a grid network, and {\small$N=600$} mobile users regularly locate over the SBSs coverage regions. We randomly create {\small$800$} transmission connects between SBSs and MUs to enable each user communicate with at least one SBS. As for SBS {\small $m$}, we set {\small $R_m=200$} and {\small$C_m=250$} by default.

2) Real request traces

In each time slot, each user submits request for service drawn from a service set, {\small$\mathcal{K}=100$}. We adopt the same service requirements in~\cite{KP}, and map services to three actual service, {\em i.e.,} Video streaming (VS), Augmented Reality (AR) and Network Gaming (NG). VS wants requirement of storage and bandwidth within {\small $\left[ 1,10 \right]$} and {\small $\left[ 1,25 \right]$}, respectively. The resource requirements of AR are set within {\small $\left[ 2,20 \right]$} and {\small $\left[ 0.2,2 \right]$}, respectively. The demand storage and bandwidth of NG are set within {\small $\left[ 5,40 \right]$} and {\small $\left[ 1,25 \right]$}, respectively. The service parameters are in accordance with real service specifications~\cite{KP}.

3) Online algorithm description

As for the online algorithm,  we set the default size of look-ahead window to be {\small $5$} and define the number of time-slot as {\small $300$} . In terms of unit storage cost of a SBS, we adopt the similar pattern in~\cite{jiaolei}, where storage price is inversely proportion to the occupied capacity. The transmission efficiency parameter {\small$d_{m,n}$} is randomly distributed in {\small $\left[ 1,5 \right]$} due to short distance and low delay. We set the dynamic placing cost {\small$b_{m,k}$} randomly within {\small $\left[ 5,10 \right]$}.

\subsection{Comparison with offline algorithm}

We introduce the algorithm competitive ratio, which can be obtained through its total cost divided by the offline optimum that can be optimally solved by the MOSEK Optimizer.

\vspace{-2mm}
\begin{figure}[htbp] 
	\centering
	\includegraphics[width=0.43\textwidth]{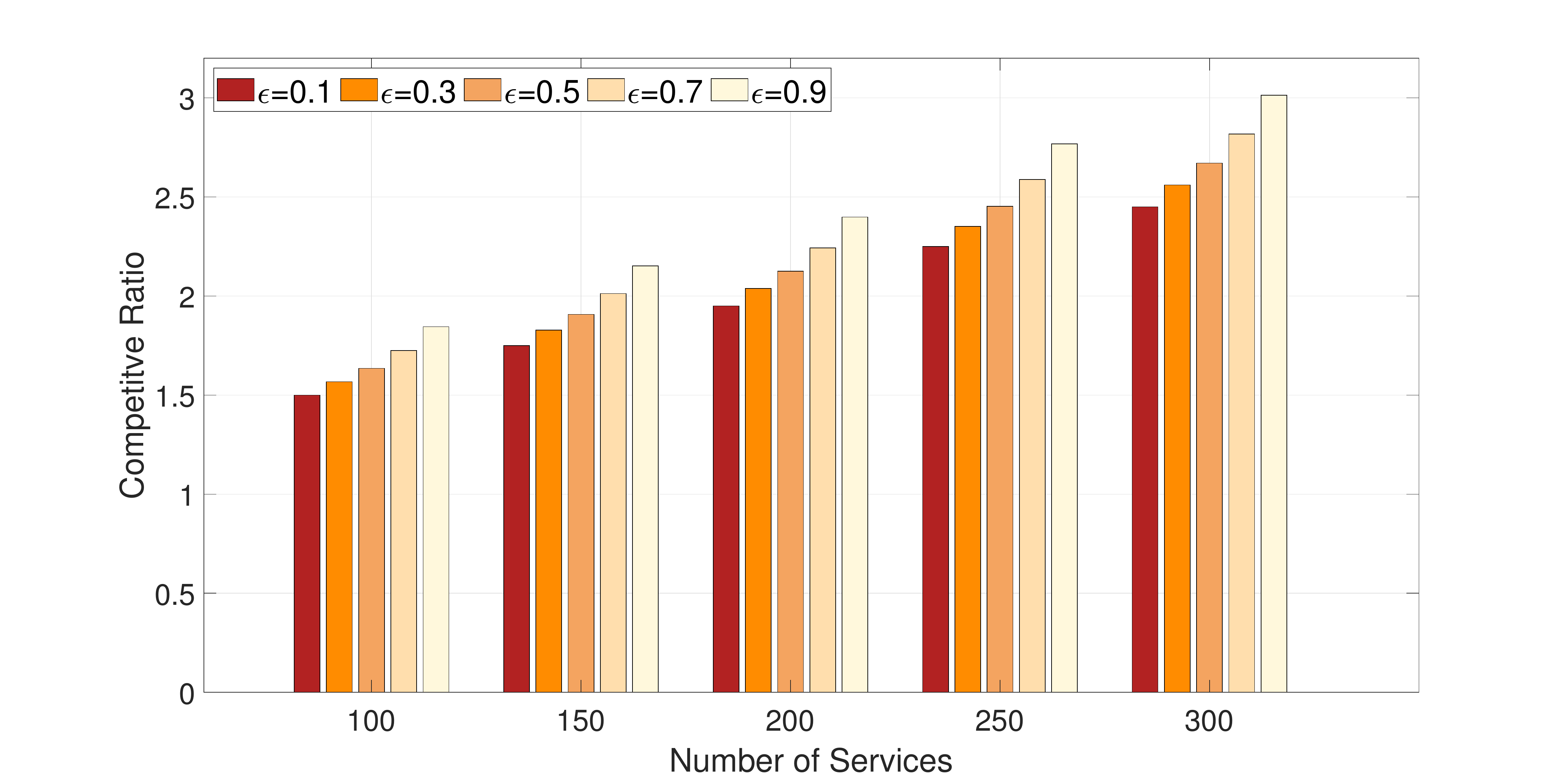}	
	\vspace{-1mm}
	\caption{Competitive ratio with different $\mathcal{K}$ and $\varepsilon$.}
\end{figure}
\vspace{-2mm}

\vspace{-2mm}
\begin{figure}[htbp] 
	\centering
	\includegraphics[width=0.43\textwidth]{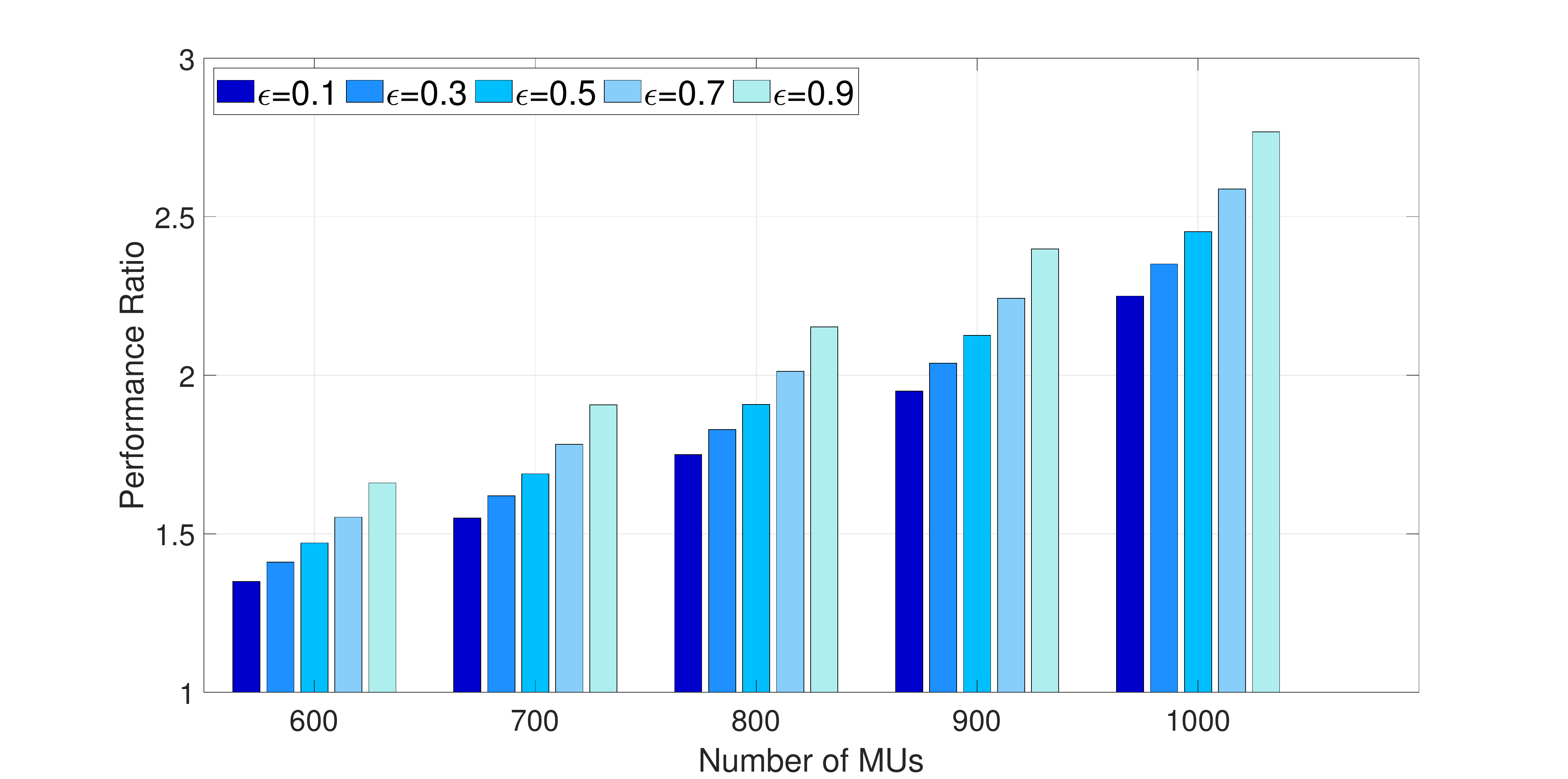}	
	\vspace{-1mm}
	\caption{Competitive ratio with different $\mathcal{N}$ and $\varepsilon$.}
\end{figure}
\vspace{-2mm}

\vspace{-2mm}
\begin{figure}[htbp] 
	\centering
	\includegraphics[width=0.43\textwidth]{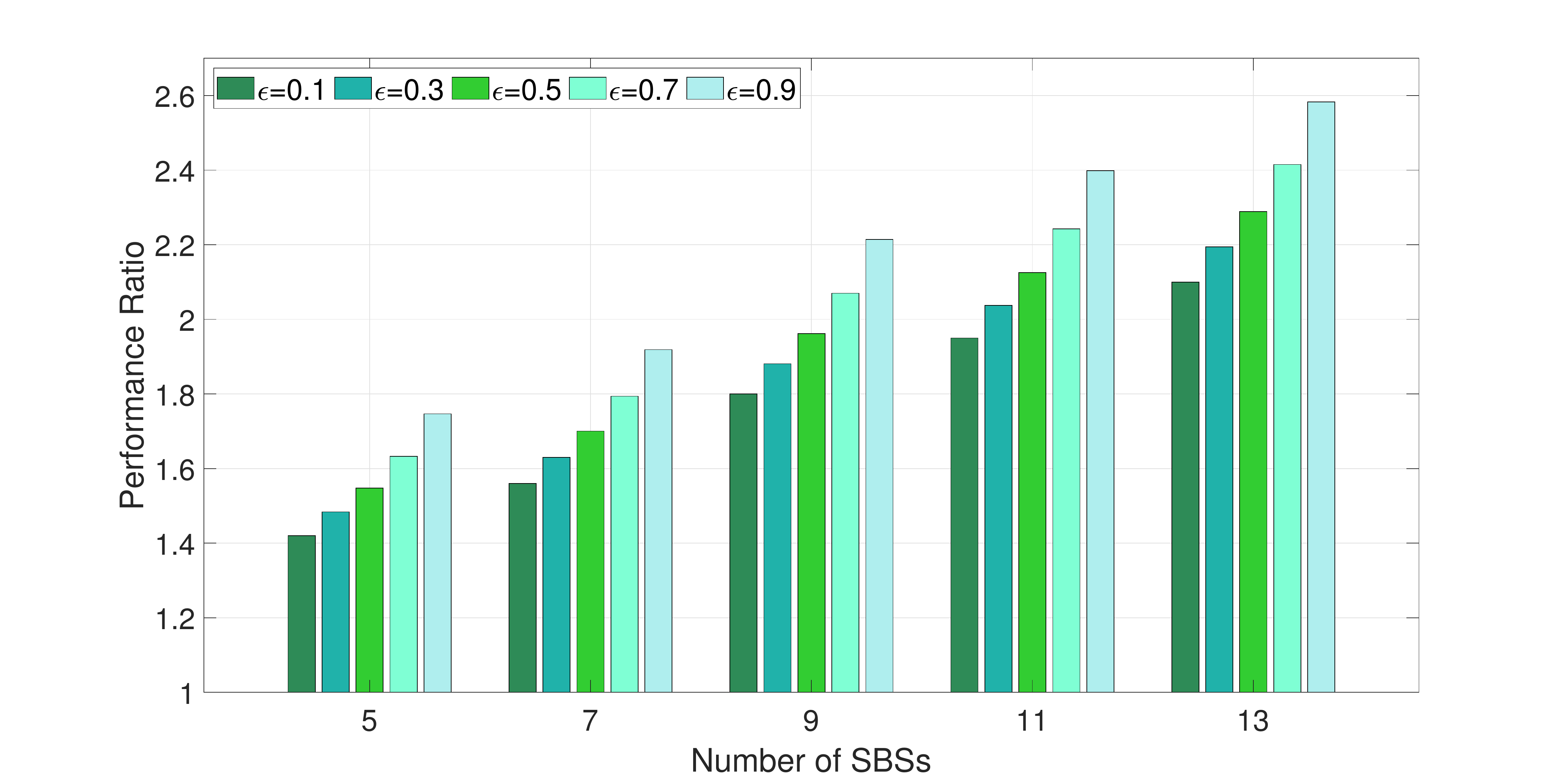}	
	\vspace{-1mm}
	\caption{Competitive ratio with different $\mathcal{M}$ and $\varepsilon$.}
\end{figure}
\vspace{-2mm}

Fig.3-Fig.5 exhibit that how the performance of online algorithm change. Globally, the increasing number of total services, MUs and SBSs leads to a larger competitive ratio, which is inline with the theoretical analysis. As we observe, Fig.3-Fig.5 also depict the impact of {\small$\varepsilon$} on our performance. In the following evaluation, we set the default value of {\small$\varepsilon$} to be {\small$0.3$} instead of the optimum {\small$\varepsilon$} to optimize the theoretical competitive ratio.

\subsection{Comparison with other schemes}

In this context, we compare with four algorithms in the latest related works.

Regularization-based algorithm (REG)~\cite{jiaolei}: it leverages regularization method to make the joint placing and routing decision, in order to minimize the overall cost;

Primal-dual algorithm (PD)~\cite{Zeng}: it resorts to primal-dual method to minimize the total cost;

Greedy algorithm (GA): it greedily make the joint decision to minimize the holistic cost;

Cache-only algorithm (CA)~\cite{CA}: it places services on SBSs until all storage capacity are full and tries to minimize the service placing cost.

\vspace{-2mm}
\begin{figure}[htbp] 
	\centering
	\includegraphics[width=0.43\textwidth]{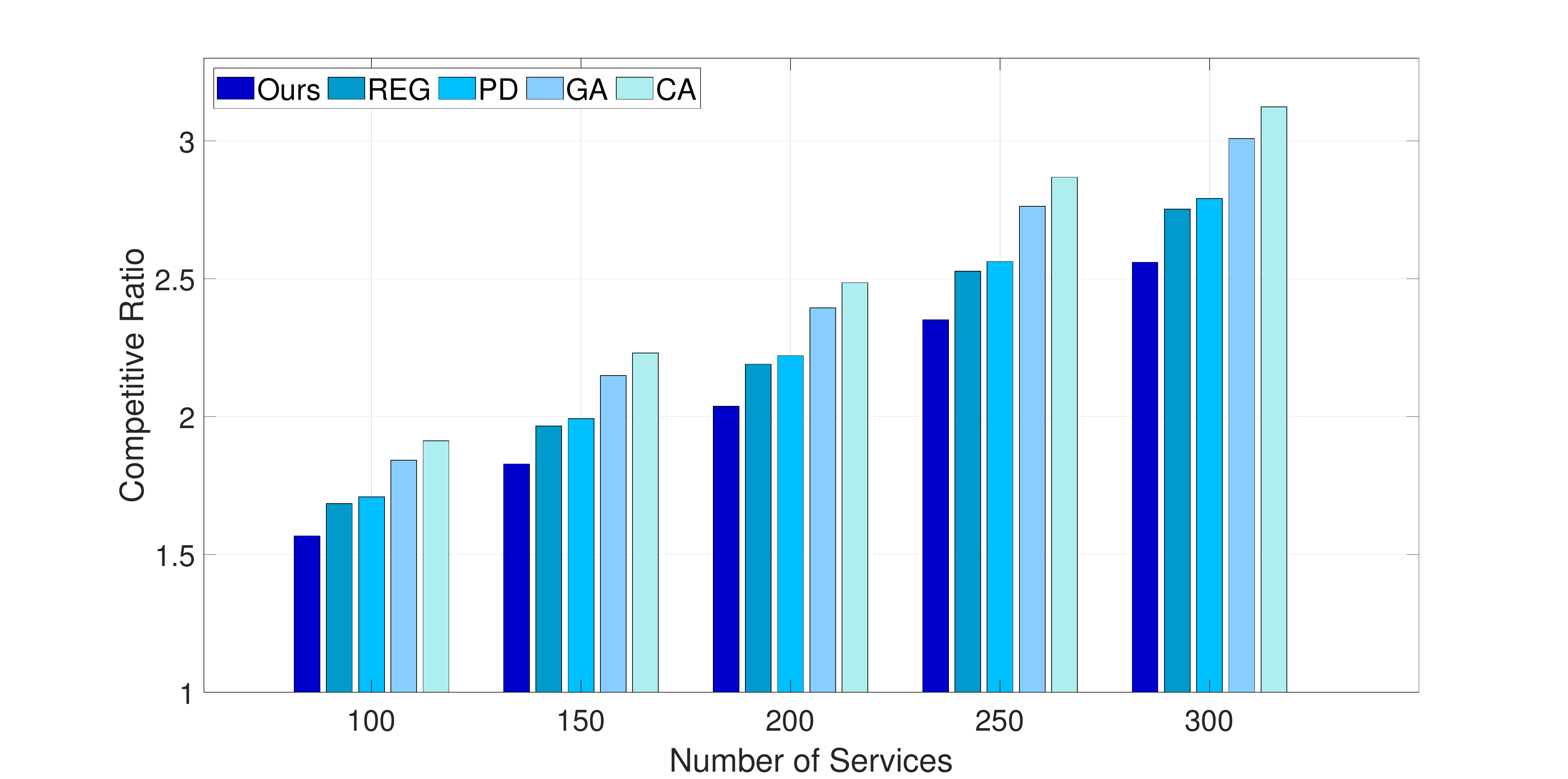}	
	\vspace{-1mm}
	\caption{Performance ratio of different algorithms.}
\end{figure}
\vspace{-2mm}

\vspace{-2mm}
\begin{figure}[htbp] 
	\centering
	\includegraphics[width=0.43\textwidth]{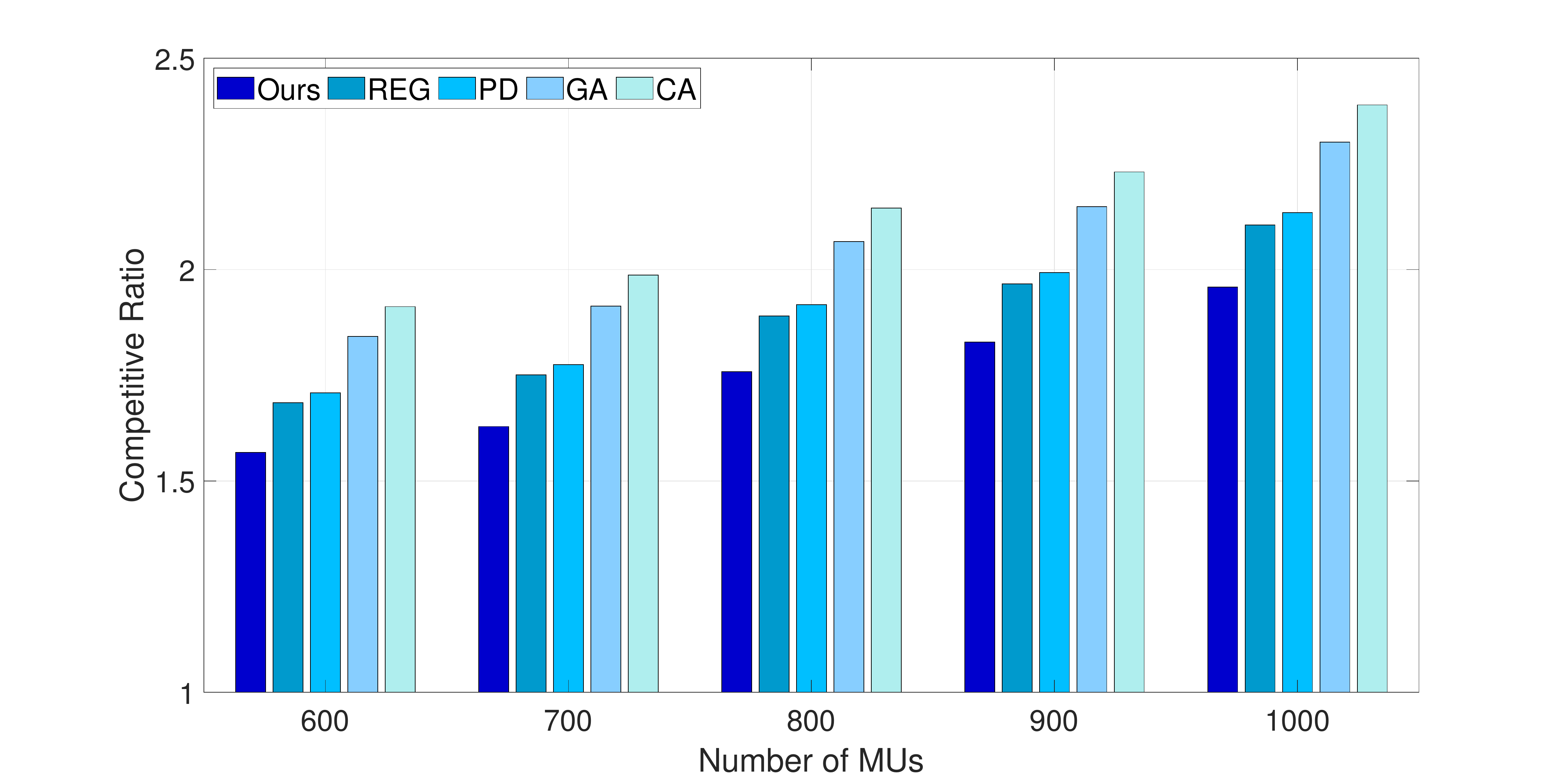}	
	\vspace{-1mm}
	\caption{Performance ratio of different algorithms.}
\end{figure}
\vspace{-2mm}

\vspace{-2mm}
\begin{figure}[htbp] 
	\centering
	\includegraphics[width=0.43\textwidth]{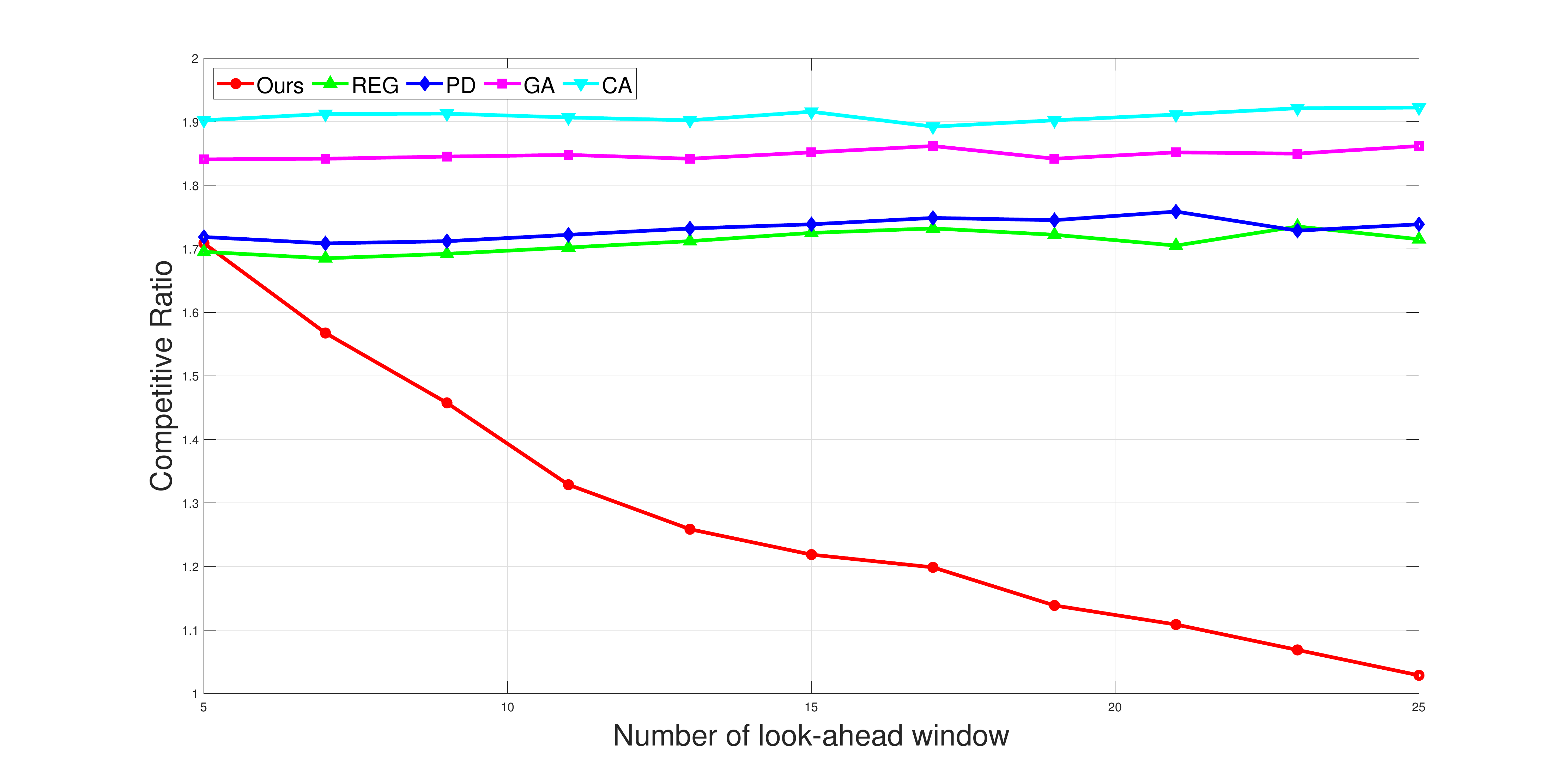}	
	\vspace{-1mm}
	\caption{Competitive ratio.}
\end{figure}

The results of Fig.6-Fig.8 present how the performance change with the increase of services, MUs, and look-ahead window size. As expected, the performance of Cache-only algorithm is the worst, since it stores as many services as possible, neglecting the transmission cost and dynamic placing cost. Greedy algorithm performs better than Cache algorithm since it considers more system cost. As we observe, the performance of REG algorithm and PD algorithm are both better than Greedy algorithm. In the contrast, without preparation for future information, there is a considerable gap between PD and ours or REG and ours. Knowing more future information helps the system making more wise decisions. Therefore, our online algorithm has a superior performance.

\begin{figure}[htbp] 
	\centering
	\includegraphics[width=0.43\textwidth]{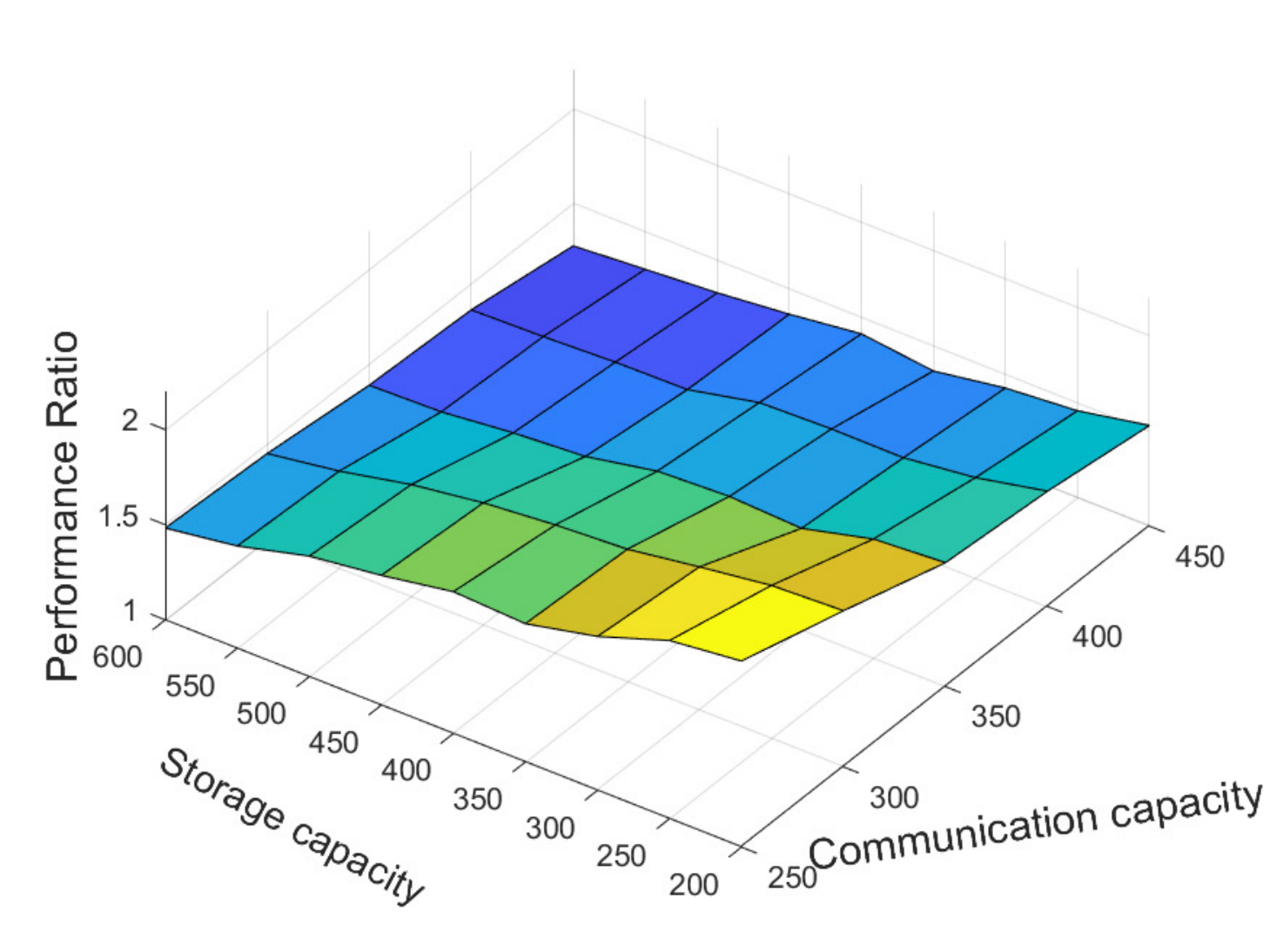}	
	\vspace{-1mm}
	\caption{Competitive ratio with different capacity}
\end{figure}

Moreover, we verify our online algorithm is compatible with various system settings. We fix these values remain unchanged but change the limited capacity. Fig.9 exhibits when the storage and communication capacity increases, our algorithm persistently has a better performance since it makes full use of system resources.
\section{Conclusion} \label{sec:conclude}

In this context, we study the joint service placing and request scheduling problem in MEC networks, and developed an efficient online framework, aiming to minimize holistic operational cost. Our online algorithm employs the competitive OCO technique and rounding method to solve the cost-minimization problem. The theoretical analysis and simulations have corroborated the efficiency of our proposed online algorithm. We confirmed that our algorithm has a superior performance over alternative benchmarks through large-scale evaluations. Next, we will further concentrate on typical real-world scenarios and design more efficient online algorithm which can accommodate more flexible user requests.

\newpage
\bibliographystyle{IEEEtran}
\bibliography{IEEEabrv,main}

\begin{thebibliography}{10}
\providecommand{\url}[1]{#1}
\csname url@samestyle\endcsname
\providecommand{\newblock}{\relax}
\providecommand{\bibinfo}[2]{#2}
\providecommand{\BIBentrySTDinterwordspacing}{\spaceskip=0pt\relax}
\providecommand{\BIBentryALTinterwordstretchfactor}{4}
\providecommand{\BIBentryALTinterwordspacing}{\spaceskip=\fontdimen2\font plus
\BIBentryALTinterwordstretchfactor\fontdimen3\font minus
  \fontdimen4\font\relax}
\providecommand{\BIBforeignlanguage}[2]{{%
\expandafter\ifx\csname l@#1\endcsname\relax
\typeout{** WARNING: IEEEtran.bst: No hyphenation pattern has been}%
\typeout{** loaded for the language `#1'. Using the pattern for}%
\typeout{** the default language instead.}%
\else
\language=\csname l@#1\endcsname
\fi
#2}}
\providecommand{\BIBdecl}{\relax}
\BIBdecl

\bibitem{MEC}
P.~{Mach} and Z.~{Becvar}, ``Mobile edge computing: A survey on architecture
  and computation offloading,'' \emph{IEEE Communications Surveys Tutorials},
  vol.~19, no.~3, pp. 1628--1656, 2017.

\bibitem{MEC-2}
Y.~Mao, C.~You, J.~Zhang, K.~Huang, and K.~B. Letaief, ``A survey on mobile
  edge computing: The communication perspective,'' \emph{IEEE Communications
  Surveys Tutorials}, vol.~19, no.~4, pp. 2322--2358, 2017.

\bibitem{5G}
X.~Ge, S.~Tu, G.~Mao, C.-X. Wang, and T.~Han, ``5g ultra-dense cellular
  networks,'' \emph{IEEE Wireless Communications}, vol.~23, no.~1, pp. 72--79,
  2016.

\bibitem{latency-2}
T.~X. {Tran} and D.~{Pompili}, ``Adaptive bitrate video caching and processing
  in mobile-edge computing networks,'' \emph{IEEE Transactions on Mobile
  Computing}, vol.~18, no.~9, pp. 1965--1978, 2019.

\bibitem{latency-1}
M.~{Dehghan}, B.~{Jiang}, A.~{Seetharam}, T.~{He}, T.~{Salonidis}, J.~{Kurose},
  D.~{Towsley}, and R.~{Sitaraman}, ``On the complexity of optimal request
  routing and content caching in heterogeneous cache networks,'' \emph{IEEE/ACM
  Transactions on Networking}, vol.~25, no.~3, pp. 1635--1648, 2017.

\bibitem{latency-3}
J.~{Li}, T.~{Khoa Phan}, W.~{Koong Chai}, D.~{Tuncer}, G.~{Pavlou},
  D.~{Griffin}, and M.~{Rio}, ``Dr-cache: Distributed resilient caching with
  latency guarantees,'' in \emph{IEEE INFOCOM 2018 - IEEE Conference on
  Computer Communications}, 2018, pp. 441--449.

\bibitem{Xu}
Z.~{Xu}, W.~{Liang}, W.~{Xu}, M.~{Jia}, and S.~{Guo}, ``Efficient algorithms
  for capacitated cloudlet placements,'' \emph{IEEE Transactions on Parallel
  and Distributed Systems}, vol.~27, no.~10, pp. 2866--2880, 2016.

\bibitem{2019Winning}
B.~{Gao}, Z.~{Zhou}, F.~{Liu}, and F.~{Xu}, ``Winning at the starting line:
  Joint network selection and service placement for mobile edge computing,'' in
  \emph{IEEE INFOCOM 2019 - IEEE Conference on Computer Communications}, 2019,
  pp. 1459--1467.

\bibitem{XuJ}
J.~{Xu}, L.~{Chen}, and P.~{Zhou}, ``Joint service caching and task offloading
  for mobile edge computing in dense networks,'' in \emph{IEEE INFOCOM 2018 -
  IEEE Conference on Computer Communications}, 2018, pp. 207--215.

\bibitem{KP}
K.~{Poularakis}, J.~{Llorca}, A.~M. {Tulino}, I.~{Taylor}, and L.~{Tassiulas},
  ``Service placement and request routing in mec networks with storage,
  computation, and communication constraints,'' \emph{IEEE/ACM Transactions on
  Networking}, vol.~28, no.~3, pp. 1047--1060, 2020.

\bibitem{cache-hit-1}
T.~{He}, H.~{Khamfroush}, S.~{Wang}, T.~{La Porta}, and S.~{Stein}, ``It's hard
  to share: Joint service placement and request scheduling in edge clouds with
  sharable and non-sharable resources,'' in \emph{2018 IEEE 38th International
  Conference on Distributed Computing Systems (ICDCS)}, 2018, pp. 365--375.

\bibitem{KP-2}
K.~{Poularakis}, G.~{Iosifidis}, and L.~{Tassiulas}, ``Approximation algorithms
  for mobile data caching in small cell networks,'' \emph{IEEE Transactions on
  Communications}, vol.~62, no.~10, pp. 3665--3677, 2014.

\bibitem{Taleb}
T.~{Taleb}, P.~A. {Frangoudis}, I.~{Benkacem}, and A.~{Ksentini}, ``Cdn slicing
  over a multi-domain edge cloud,'' \emph{IEEE Transactions on Mobile
  Computing}, vol.~19, no.~9, pp. 2010--2027, 2020.

\bibitem{Yang}
L.~{Yang}, J.~{Cao}, G.~{Liang}, and X.~{Han}, ``Cost aware service placement
  and load dispatching in mobile cloud systems,'' \emph{IEEE Transactions on
  Computers}, vol.~65, no.~5, pp. 1440--1452, 2016.

\bibitem{Ceselli}
A.~{Ceselli}, M.~{Premoli}, and S.~{Secci}, ``Mobile edge cloud network design
  optimization,'' \emph{IEEE/ACM Transactions on Networking}, vol.~25, no.~3,
  pp. 1818--1831, 2017.

\bibitem{2020Intelligent}
F.~Wang, F.~Wang, J.~Liu, R.~Shea, and L.~Sun, ``Intelligent video caching at
  network edge: A multi-agent deep reinforcement learning approach,'' in
  \emph{IEEE INFOCOM 2020 - IEEE Conference on Computer Communications}, 2020.

\bibitem{Zeng}
Y.~{Zeng}, Y.~{Huang}, Z.~{Liu}, and Y.~{Yang}, ``Online distributed edge
  caching for mobile data offloading in 5g networks,'' in \emph{2020 IEEE/ACM
  28th International Symposium on Quality of Service (IWQoS)}, 2020, pp. 1--10.

\bibitem{Zhao}
T.~{Zhao}, I.~. {Hou}, S.~{Wang}, and K.~{Chan}, ``Red/led: An asymptotically
  optimal and scalable online algorithm for service caching at the edge,''
  \emph{IEEE Journal on Selected Areas in Communications}, vol.~36, no.~8, pp.
  1857--1870, 2018.

\bibitem{OCO}
X.~L. M.~Shi and L.~Jiao, ``Combining regularization with look-ahead for
  competitive online convex optimization,'' purdue University, Tech. Rep.,
  2021. Available at https://engineering.purdue.edu/\%7elinx/papers.html.

\bibitem{1980Complexity}
G.~Gens and E.~Levner, ``Complexity of approximation algorithms for
  combinatorial problems: a survey,'' \emph{ACM SIGACT News}, vol.~12, no.~3,
  pp. 52--65, 1980.

\bibitem{zhouzhi}
Z.~{Zhou}, Q.~{Wu}, and X.~{Chen}, ``Online orchestration of cross-edge service
  function chaining for cost-efficient edge computing,'' \emph{IEEE Journal on
  Selected Areas in Communications}, vol.~37, no.~8, pp. 1866--1880, 2019.

\bibitem{regularization}
N.~Buchbinder, S.~Chen, and J.~S. Naor, \emph{Competitive Analysis via
  Regularization}, 2014.

\bibitem{Interior}
A.~Wächter and L.~T. Biegler, ``On the implementation of a primal-dual
  interior point filter line search algorithm for large-scale nonlinear
  programming,'' \emph{Mathematical Programming}, vol. 106, no.~1, pp. 25--57,
  2006.

\bibitem{primal-dual}
N.~Buchbinder, S.~Chen, J.~Naor, and O.~Shamir, ``Unified algorithms for online
  learning and competitive analysis,'' \emph{Mathematics of Operations
  Research}, vol.~41, pp. 5.1--5.18, 2016.

\bibitem{Convex}
S.~Boyd and L.~Vandenberghe, \emph{Convex Optimization}, 2004.

\bibitem{independent}
P.~Raghavan and C.~D. Tompson, \emph{Randomized rounding: a technique for
  provably good algorithms and algorithmic proofs}.\hskip 1em plus 0.5em minus
  0.4em\relax Springer-Verlag New York, Inc., 1987.

\bibitem{Rounding}
R.~Gandhi, S.~Khuller, S.~Parthasarathy, and A.~Srinivasan, ``Dependent
  rounding and its applications to approximation algorithms,'' \emph{Journal of
  the ACM}, vol.~53, no.~3, pp. 324--360, 2006.

\bibitem{jiaolei}
L.~{Pu}, L.~{Jiao}, X.~{Chen}, L.~{Wang}, Q.~{Xie}, and J.~{Xu}, ``Online
  resource allocation, content placement and request routing for cost-efficient
  edge caching in cloud radio access networks,'' \emph{IEEE Journal on Selected
  Areas in Communications}, vol.~36, no.~8, pp. 1751--1767, 2018.

\bibitem{CA}
A.~Gharaibeh, A.~Khreishah, B.~Ji, and M.~Ayyash, ``A provably efficient online
  collaborative caching algorithm for multicell-coordinated systems,''
  \emph{IEEE Transactions on Mobile Computing}, vol.~15, no.~8, pp. 1863--1876,
  2016.

\end{thebibliography}

\begin{IEEEbiography}[{\includegraphics[width=1in,height=1.25in,clip,keepaspectratio]{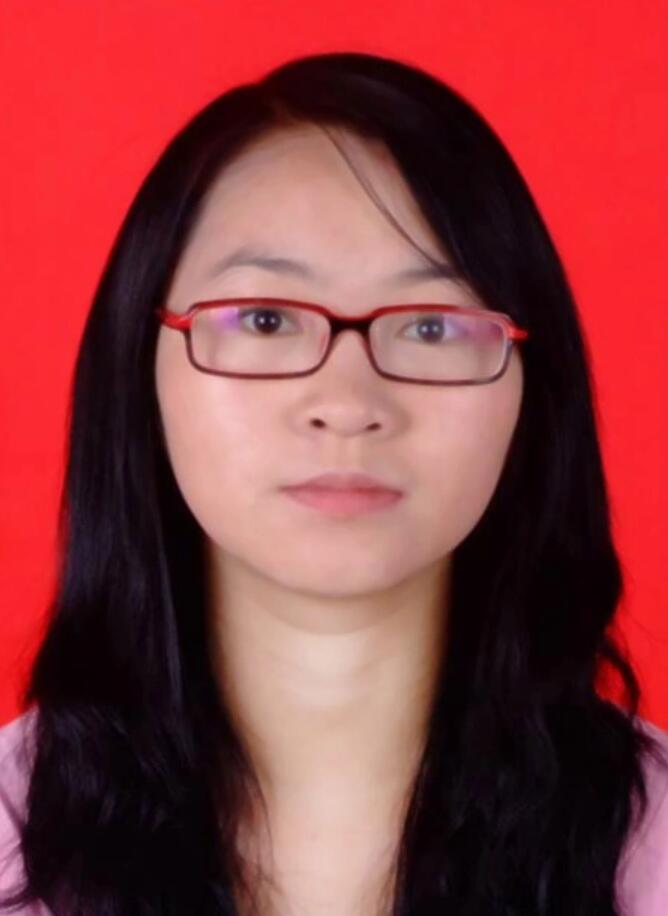}}]{Lina Su}
received her M.S. degree in 2011 from School of Computer Science, Chongqing University of Posts and Telecommunications, China. Since September, 2019, she has been  a Ph.D student in School of Computer Science, Wuhan University. Her research includes online learning, online optimization, mobile edge computing and Internet of Things.
\end{IEEEbiography}

\begin{IEEEbiography}[{\includegraphics[width=1in,height=1.25in,clip,keepaspectratio]{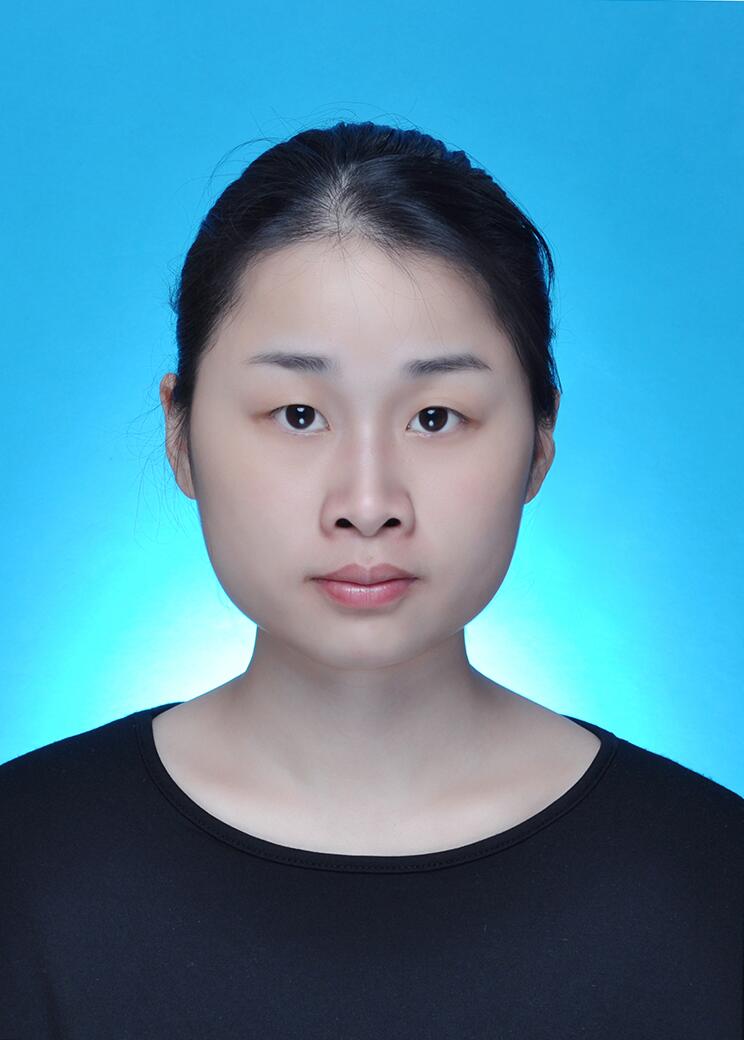}}]{Ne Wang}
received her B.E. and M.S. degrees in 2016 and 2019 from School of Computer Science, Wuhan University Of Technology, China. Since September, 2019, she has been a Ph.D student in School of Computer Science, Wuhan University, China. Her research interests is in the areas of cloud computing, machine learning optimization, and online scheduling.
\end{IEEEbiography}

\begin{IEEEbiography}[{\includegraphics[width=1in,height=1.25in,clip,keepaspectratio]{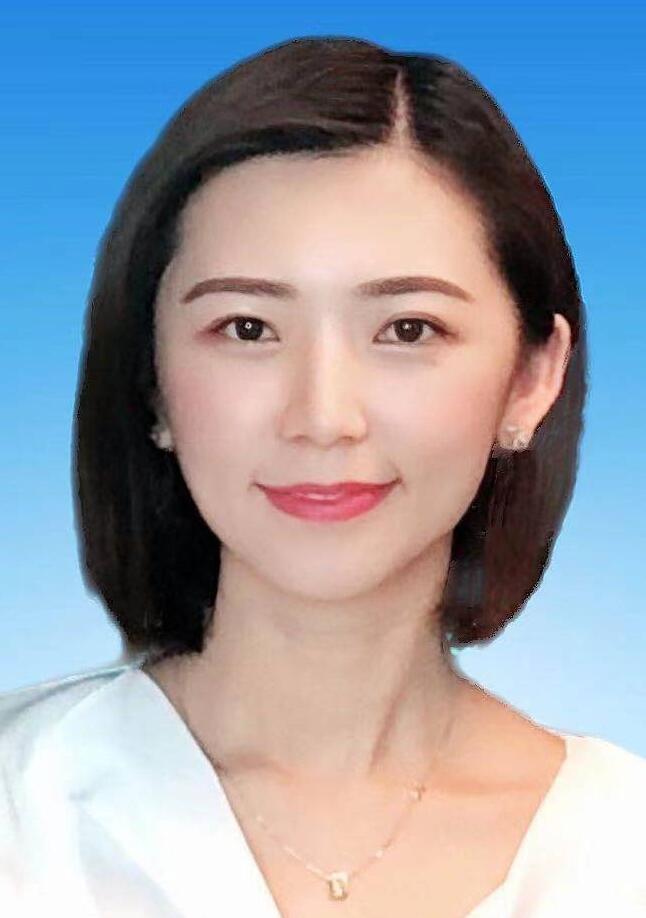}}]{Ruiting Zhou} has been an Associate Professor in the School of Cyber Science and Engineering at Wuhan University since June 2018. She received her Ph.D. degree in 2018 from the Department of Computer Science, University of Calgary, Canada. Her research interests include cloud computing, machine learning and mobile network optimization. She has published research papers in top-tier computer science conferences and journals, including IEEE INFOCOM, ACM Mobihoc, IEEE/ACM TON, IEEE JSAC, IEEE TMC. She serves as the TPC chair for INFOCOM workshop-ICCN2019/2020/2021. She also serves as a reviewer for international conferences and journals  such us IEEE/ACM IWQoS, IEEE Globecom, IEEE JSAC, IEEE TON, IEEE TMC, IEEE TCC and IEEE TWC.
\end{IEEEbiography}

\begin{IEEEbiography}[{\includegraphics[width=1in,height=1.25in,clip,keepaspectratio]{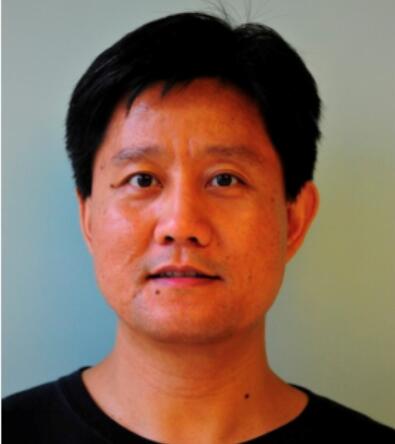}}]{Zongpeng Li}
received the BE degree in computer science from Tsinghua University, in 1999, and the PhD degree from the University of Toronto, in 2005. He has been with the University of Calgary and then Wuhan University. His research interests are in computer networks and cloud computing. He was named an Edward S.Rogers Sr. Scholar, in 2004, won the Alberta Ingenuity New Faculty Award, in 2007, and was nominated for the Alfred P. Sloan Research Fellow, in 2007. He co-authored papers that received Best Paper Awards at the following conferences: PAM 2008, HotPOST 2012, and ACM e-Energy 2016. He received the Department Excellence Award from the Department of Computer Science, University of Calgary, the Outstanding Young Computer Science Researcher Prize from the Canadian Association of Computer Science, and the Research Excellence Award from the Faculty of Science, University of Calgary. He is a senior member of the IEEE.
\end{IEEEbiography}

\newpage

{\bf Appendix A }

{\bf Proof of Lemma 1}. 

To prove Lemma 1, we start with proving one episode from {\small$t^{(\pi)}$} to {\small$t^{(\pi)}+L$}. The proof of other episodes are similar.

Firstly, based on 
\vspace{-3mm}
\begin{equation}
    \label{eq:KKT}
      \text{\bf KKT Optimality Conditions}
\end{equation}

of {\small $P_{ORA}^{(\pi)}$}, we derive

{\small
\begin{align}
 l_{m,k} -   \sum_{n}\theta_{m,n,k}^{(\pi)}(t^{(\pi)}) +r_k\rho_m^{(\pi)}(t^{(\pi)}) - \beta_{m,k}^{(\pi)}(t^{(\pi)}+1) +\frac{b_{m,k}}{\eta}\ln\left(\frac{1+\frac{\varepsilon}{\mathcal{M}\mathcal{K}}}{ x_{m,k}^{(\pi)}(t^{(\pi)}-1)+\frac{\varepsilon}{\mathcal{M}\mathcal{K}}}\right)\ge 0,\forall m,k, \tag{\ref{eq:KKT}a}
\end{align}}
\vspace{-4mm}
\vspace{-4mm}
{\small
\begin{align}
  l_{m,k} -  \sum_{n}\theta_{m,n,k}^{(\pi)}(t) +  r_k\rho_m^{(\pi)}(t) + \beta_{m,k}^{(\pi)}(t)- \beta_{m,k}^{(\pi)}(t+1) \ge 0, \forall m,k,t \in [t^{(\pi)}+1,t^{(\pi)}+L-1],\tag{\ref{eq:KKT}b}
\end{align}}
\vspace{-4mm}
\vspace{-4mm}
{\small
\begin{align}
l_{m,k} -   \sum_{n}\theta_{m,n,k}^{(\pi)}(t^{(\pi)}+L) +r_k\rho_m^{(\pi)}(t^{(\pi)}+L) + \beta_{m,k}^{(\pi)}(t^{(\pi)}+L)  - \frac{b_{m,k}}{\eta}\ln\left(\frac{1+\frac{\varepsilon}{\mathcal{M}\mathcal{K}}}{ x_{m,k}^{(\pi)}(t^{(\pi)}+L)+\frac{\varepsilon}{\mathcal{M}\mathcal{K}}}\right)\ge 0,
 \forall m,k, \tag{\ref{eq:KKT}c}
\end{align}}
\vspace{-4mm}
\vspace{-4mm}
{\small
\begin{align}
d_{m,n}\lambda_{n,k}(t)-\alpha_{n,k}^{(\pi)}(t)+ \theta_{m,n,k}^{(\pi)}(t)+c_k\lambda_{n,k}(s)\mu_m^{(\pi)}(t)\ge 0, \forall m,n,k,t \in [t^{(\pi)}+1,t^{(\pi)}+L], \tag{\ref{eq:KKT}d}
\end{align}}
\vspace{-4mm}
\vspace{-4mm}
{\small
\begin{align}
 b_{m,k} \geq \beta_{m,k}^{(\pi)}(t), \forall m,n,k,t \in [t^{(\pi)}+1,t^{(\pi)}+L].\tag{\ref{eq:KKT}e}
\end{align}}
\vspace{-4mm}

Therefore, constraint (\ref{eq:max}a) from {\small $t^{(\pi)}+1$} to {\small $t^{(\pi)}+L-1$} is satisfied. And constraint (\ref{eq:max}b), constraint (\ref{eq:max}c), constraint {\small $\theta_{m,n,k}(t)\ge 0 , \alpha_{n,k}(t)\ge 0, \rho_m(t) \ge 0, \mu_m(t) \ge 0$} from {\small $t^{(\pi)}$} to {\small $t^{(\pi)}+L$}, and constraint {\small $\beta_{m,k}(t)\ge 0$} from {\small $t^{(\pi)}+1$} to {\small $t^{(\pi)}+L$} are satisfied. In addition, based on (\ref{eq:beta_1}), we obtain {\small $\beta_{m,k}^{(\pi)}(t^{(\pi)})=  \frac{b_{m,k}}{\eta}\ln(\frac{1+\frac{\varepsilon}{\mathcal{M}\mathcal{K}}}{x_{m,k}^{(\pi)}(t^{(\pi)}-1)+\frac{\varepsilon}{\mathcal{M}\mathcal{K}}})$} and  {\small $\beta_{m,k}^{(\pi)}(t^{(\pi)}+L)=  \frac{b_{m,k}}{\eta}\ln(\frac{1+\frac{\varepsilon}{\mathcal{M}\mathcal{K}}}{x_{m,k}^{(\pi)}(t^{(\pi)}+L)+\frac{\varepsilon}{\mathcal{M}\mathcal{K}}})$}. According to (\ref{eq:KKT}a) and (\ref{eq:KKT}c), we can conduct that constraint (\ref{eq:max}a) at time {\small $t^{(\pi)}$} and {\small $t^{(\pi)}+L$}, and constraint {\small $\beta_{m,k}(t)\ge 0$} at time {\small $t^{(\pi)}$} is satisfied. \qed

\vspace{3mm}
{\bf Appendix B }

{\bf Proof of Lemma 2}.  

As Alg.1 described, if {\small $t^{(\pi)} \leq 0$}, the third term in the objective function should be removed, and if {\small $t^{(\pi)}+L \geq T$}, the fifth term should be removed. In the following, we reformulate Lemma 2 in a precise manner by partitioning the time domain into three parts {\small $t_b^{(\pi)} \in [-(L+1),0]$}, {\small $t_m^{(\pi)} \in [1,T-L-1]$} and {\small $t_e^{(\pi)} \in [T-L,T]$}, {\em i.e.,} the beginning, middle and end segment.

{\bf\em Lemma 6}. For each {\small $ORA^{(\pi)}$}, we derive

{\small
\begin{equation*}
\begin{aligned}
	C^{(\pi)}(1:t_b^{(\pi)}+L) =  D^{(\pi)}(1:t_b^{(\pi)}+L)+\sum\limits_{m}\sum\limits_{k}{\psi}_{m,k}^{(\pi)}(t_b^{(\pi)}),
\end{aligned}
\end{equation*}}	
\vspace{-2mm}
\vspace{-2mm}
{\small
\begin{equation*}
\begin{aligned}
	C^{(\pi)}(t_m^{(\pi)}:t_m^{(\pi)}+L) =  D^{(\pi)}(t_m^{(\pi)}:t_m^{(\pi)}+L) +\sum\limits_{m}\sum\limits_{k}{\Omega}_{m,k}^{(\pi)}(t_m^{(\pi)}) 
  +\sum\limits_{m}\sum\limits_{k}{\phi}_{m,k}^{(\pi)}(t_m^{(\pi)})+\sum\limits_{m}\sum\limits_{k}{\psi}_{m,k}^{(\pi)}(t_m^{(\pi)}),
\end{aligned}
\end{equation*}}	
\vspace{-2mm}
\vspace{-2mm}
{\small
\begin{equation*}
\begin{aligned}
	C^{(\pi)}(t_e^{(\pi)}:T) =  D^{(\pi)}(t_e^{(\pi)}:T) +\sum\limits_{m}\sum\limits_{k}{\Omega}_{m,k}^{(\pi)}(t_e^{(\pi)})  +\sum\limits_{m}\sum\limits_{k}{\phi}_{m,k}^{(\pi)}(t_e^{(\pi)}),
\end{aligned}
\end{equation*}}	
\vspace{-6mm}

\noindent where {\small $t_b^{(\pi)},t_m^{(\pi)},t_e^{(\pi)}=\pi+(L+1)v;v=-1,0,\dots,\lceil \frac{T}{L+1} \rceil$}, such that {\small $t_b^{(\pi)} \in [-(L+1),0]$}, {\small $t_m^{(\pi)} \in [1,T-L-1]$} and {\small $t_e^{(\pi)} \in [T-L,T]$}.

Proof.
Firstly, we prove the episode starting from time {\small $t_m^{(\pi)} \in [1,T-L-1]$}. And, we reformulate {\small $P_{ORA}^{(\pi)}$} as follows 
\vspace{-2mm}
{\small
\begin{equation}
\label{eq:min4}
\begin{aligned}
\text{minimize}	
  & \sum\limits_{t=t_m^{(\pi)}}^{t_m^{(\pi)}+L}\sum\limits_{m}\sum\limits_{k}l_{m,k}x_{m,k}(t) + \sum\limits_{t=t_m^{(\pi)}}^{t_m^{(\pi)}+L}\sum\limits_{m}\sum\limits_{n}\sum\limits_{k}d_{m,n}{\lambda_{n,k}(t)}{y_{m,n,k}(t)}\\
  & +\sum\limits_{m}\sum\limits_{k}\frac{b_{m,k}}{\eta}x_{m,k}(t_m^{(\pi)})\ln\left(\frac{1+\frac{\varepsilon}{\mathcal{M}\mathcal{K}}}{x_{m,k}^{(\pi)}(t_m^{(\pi)}-1)+\frac{\varepsilon}{\mathcal{M}\mathcal{K}}}\right)+\sum\limits_{t=t_m^{(\pi)}+1}^{t_m^{(\pi)}+L}\sum\limits_{m}\sum\limits_{k}b_{m,k}z_{m,k}(t)\\
  &+\sum\limits_{m}\sum\limits_{k}\frac{b_{m,k}}{\eta} \left[\left(x_{m,k}(t_m^{(\pi)}+L)+\frac{\varepsilon}{\mathcal{M}\mathcal{K}}\right) \cdot \ln\left(\frac{x_{m,k}(t_m^{(\pi)}+L)+\frac{\varepsilon}{\mathcal{M}\mathcal{K}}}{1+\frac{\varepsilon}{\mathcal{M}\mathcal{K}}}\right)-x_{m,k}(t_m^{(\pi)}+L)\right]
\end{aligned}
\end{equation}}
\vspace{-2mm}
 
\vspace{-2mm}
{\small
subject to: 
\begin{align}
 y_{m,n,k}(t)\leq x_{m,k}(t),  \forall m,n,k,t \in [t_m^{(\pi)},t_m^{(\pi)}+L], \tag{\ref{eq:min4}a}\\
\sum_{m}y_{m,n,k}(t)\geq 1,\forall n,k,t \in [t_m^{(\pi)},t_m^{(\pi)}+L], \tag{\ref{eq:min4}b}\\
	z_{m,k}(t) \geq x_{m,k}(t)-x_{m,k}(t-1),\forall m,k,t  \in [t_m^{(\pi)}+1,t_m^{(\pi)}+L],
\tag{\ref{eq:min4}c }\\
\sum_{k}r_kx_{m,k}(t)\leq R_m,\forall m,t\in [t_m^{(\pi)},t_m^{(\pi)}+L], \tag{\ref{eq:min4}d}\\
\sum_{k}\sum_{n}y_{m,n,k}(t)\lambda_{n,k}(t)c_k\leq C_m,\forall m,t\in [t_m^{(\pi)},t_m^{(\pi)}+L], \tag{\ref{eq:min4}e}\\
y_{m,n,k}(t)\in \left[ 0,1 \right],\forall m,n,k,t\in [t_m^{(\pi)},t_m^{(\pi)}+L], \tag{\ref{eq:min4}f}\\
  x_{m,k}(t)\in\left[ 0,1 \right],\forall m,k,t\in [t_m^{(\pi)},t_m^{(\pi)}+L], \tag{\ref{eq:min4}g}\\
  z_{m,k}(t)\in\left[ 0,1 \right],\forall m,k,t\in [t_m^{(\pi)}+1,t_m^{(\pi)}+L], \tag{\ref{eq:min4}h}
\end{align}}
\noindent where {\small $\eta= \ln (1+\frac {\mathcal{M}\mathcal{K}} {\varepsilon})$}, {\small${\varepsilon} > 0$} and decision {\small $x_{m,k}^{(\pi)}(t_m^{(\pi)}-1)$} can be derived by solving {\small $ORA^{(\pi)}$} based on the previous episode from time {\small $t_m^{(\pi)}-L-1$} to {\small $t_m^{(\pi)}-1$}. 

Then, by utilizing KKT conditions~\cite{Convex} to (\ref{eq:min4}), we also acquire the Complementary slackness and Stationarity conditions, whose forms are similar to the formulation of (\ref{eq:Complementary}) and (\ref{eq:Stationarity}).
Thirdly, for each {\small $ORA^{(\pi)}$}, the overall cost from {\small $t_m^{(\pi)}$} to {\small$t_m^{(\pi)}+L$} is 

\vspace{-6mm}
{\small
\begin{equation*}
\label{eq:t_m}
\begin{aligned}
	 C^{(\pi)}(t_m^{(\pi)}:t_m^{(\pi)}+L) 
	=  &  \sum\limits_{t=t_m^{(\pi)}}^{t_m^{(\pi)}+L}\sum\limits_{m}\sum\limits_{k}l_{m,k}x^{(\pi)}_{m,k}(t)+\sum\limits_{t=t_m^{(\pi)}}^{t_m^{(\pi)}+L}\sum\limits_{m}\sum\limits_{n}\sum\limits_{k}d_{m,n}{\lambda_{n,k}(t)}{y_{m,n,k}^{(\pi)}(t)}\\
  &+\sum\limits_{t=t_m^{(\pi)}+1}^{t_m^{(\pi)}+L}\sum\limits_{m}\sum\limits_{k}b_{m,k}z^{(\pi)}_{m,k}(t)+\sum\limits_{m}\sum\limits_{k}b_{m,k}{\left[x_{m,k}^{(\pi)}(t_m^{(\pi)})-x_{m,k}^{(\pi)}(t_m^{(\pi)}-1)\right]}^{+}.
\end{aligned}
\end{equation*}}	
\vspace{-6mm}

Then, adding the Complementary slackness equations to the RHS of (\ref{eq:t_m}), we get

\vspace{-6mm}
{\small
\begin{equation}
\label{eq:middle_1}
\begin{aligned}
	 C^{(\pi)}(t_m^{(\pi)}:t_m^{(\pi)}+L) 
	=  &  \sum\limits_{t=t_m^{(\pi)}}^{t_m^{(\pi)}+L}D^{(\pi)}(t)\\
    & + \sum\limits_{m}\sum\limits_{k}x_{m,k}^{(\pi)}(t_m^{(\pi)}) \left[ l_{m,k} +r_k \rho_m^{(\pi)}(t_m^{(\pi)})-\beta_{m,k}^{(\pi)}(t_m^{(\pi)}+1) \right. \\
     & \left.-\sum_{n} \theta_{m,n,k}^{(\pi)}(t_m^{(\pi)})  +\frac{b_{m,k}}{\eta}\ln\left(\frac{ 1 +\frac{\varepsilon}{\mathcal{M}\mathcal{K}}}{ x_{m,k}^{(\pi)}(t_m^{(\pi)}-1)+\frac{\varepsilon}{\mathcal{M}\mathcal{K}}}\right)\right] \\  
    & + \sum\limits_{t=t_m^{(\pi)}+1}^{t_m^{(\pi)}+L\atop-1}\sum\limits_{m}\sum\limits_{n}\sum\limits_{k} x_{m,k}^{(\pi)}(t) \left[ l_{m,k} -\sum_{n} \theta_{m,n,k}^{(\pi)}(t)+r_k \rho_m^{(\pi)}(t)+\beta_{m,k}^{(\pi)}(t) -\beta_{m,k}^{(\pi)}(t+1)\right]\\
    & + \sum\limits_{m}\sum\limits_{k}x_{m,k}^{(\pi)}(t_m^{(\pi)}+L) \left[r_k \rho_m^{(\pi)}(t_m^{(\pi)}+L)+\beta_{m,k}^{(\pi)}(t_m^{(\pi)}+L)\right. \\
  & \left.+l_{m,k}-\sum_{n} \theta_{m,n,k}^{(\pi)}(t_m^{(\pi)}+L) -\frac{b_{m,k}}{\eta}\ln\left(\frac{ 1 +\frac{\varepsilon}{\mathcal{M}\mathcal{K}}}{ x_{m,k}^{(\pi)}(t_m^{(\pi)}+L)+\frac{\varepsilon}{\mathcal{M}\mathcal{K}}}\right)\right]\\
  & + \sum\limits_{t=t_m^{(\pi)}}^{t_m^{(\pi)}+L}\sum\limits_{m}\sum\limits_{k}y_{m,n,k}^{(\pi)}(t) \left[ \theta_{m,n,k}^{(\pi)}(t) +c_k\lambda_{n,k}(t)\mu_m^{(\pi)}(t)- \alpha_{n,k}^{(\pi)}(t)  +d_{m,n}\lambda_{n,k}(t)\right]\\ 
  & + \sum\limits_{t=t_m^{(\pi)}+1}^{t_m^{(\pi)}+L}\sum\limits_{m}\sum\limits_{k} z_{m,k}^{(\pi)}(t) \left[\beta_{m,k}^{(\pi)}(t)-b_{m,k}\right] +\sum\limits_{m}\sum\limits_{k} b_{m,k}{\left[x_{m,k}^{(\pi)}(t_m^{(\pi)})-x_{m,k}^{(\pi)}(t_m^{(\pi)}-1)\right]}^{+}\\
  & -\sum\limits_{m}\sum\limits_{k} \frac{b_{m,k}}{\eta}x_{m,k}^{(\pi)}(t_m^{(\pi)})\ln\left(\frac{1+\frac{\varepsilon}{\mathcal{M}\mathcal{K}}}{x_{m,k}^{(\pi)}(t_m^{(\pi)}-1)+\frac{\varepsilon}{\mathcal{M}\mathcal{K}}}\right)\\
  & + \sum\limits_{m}\sum\limits_{k}\frac{b_{m,k}}{\eta}x^{(\pi)}_{m,k}(t_m^{(\pi)}+L)\ln\left(\frac{1+\frac{\varepsilon}{\mathcal{M}\mathcal{K}}}{x_{m,k}^{(\pi)}(t_m^{(\pi)}+L)+\frac{\varepsilon}{\mathcal{M}\mathcal{K}}}\right),
\end{aligned}
\end{equation}}	
\vspace{-6mm}

\noindent where {\small $D^{(\pi)}(t)=\sum\limits_{n}\sum\limits_{k}{\alpha}_{n,k}^{(\pi)}(t)-\sum\limits_{m}\rho_m^{(\pi)}(t)R_m-\sum\limits_{m}\mu_m^{(\pi)}(t)C_m$}. For simplicity, we use {\small$D^{(\pi)}(t)$} in the following proof.

Utilizing the Optimality condition to (\ref{eq:middle_1}), we derive

\vspace{-2mm}
{\small
\begin{equation*}
\label{eq:middle_2}
\begin{aligned}
	 C^{(\pi)}(t_m^{(\pi)}:t_m^{(\pi)}+L) 
	=  &  \sum\limits_{t=t_m^{(\pi)}}^{t_m^{(\pi)}+L}D^{(\pi)}(t)  +\sum\limits_{m}\sum\limits_{k} b_{m,k}{\left[x_{m,k}^{(\pi)}(t_m^{(\pi)})-x_{m,k}^{(\pi)}(t_m^{(\pi)}-1)\right]}^{+}\\
  & -\sum\limits_{m}\sum\limits_{k} \frac{b_{m,k}}{\eta}x_{m,k}^{(\pi)}(t_m^{(\pi)})\ln\left(\frac{1+\frac{\varepsilon}{\mathcal{M}\mathcal{K}}}{x_{m,k}^{(\pi)}(t_m^{(\pi)}-1)+\frac{\varepsilon}{\mathcal{M}\mathcal{K}}}\right)\\
  & + \sum\limits_{m}\sum\limits_{k}\frac{b_{m,k}}{\eta}x^{(\pi)}_{m,k}(t_m^{(\pi)}+L)\ln\left(\frac{1+\frac{\varepsilon}{\mathcal{M}\mathcal{K}}}{x_{m,k}^{(\pi)}(t_m^{(\pi)}+L)+\frac{\varepsilon}{\mathcal{M}\mathcal{K}}}\right).
\end{aligned}
\end{equation*}}	
\vspace{-2mm}

Thus, considering the episodes' beginning time from {\small $t \in [1,T-L-1]$}, Lemma 6 is true. 
Similarly, we can also prove the episodes' beginning time from {\small $t \in [-(L+1),0]$} and {\small $t \in [T-L,T]$}.  \qed

\vspace{2mm}
{\bf Appendix C }

{\bf Proof of Lemma 3}.

Next, we prove (\ref{eq:tail_1}) and (\ref{eq:tail_2}) one by one.

A. Proof of (\ref{eq:tail_1}).

Firstly, let {\small$t_{l\downarrow}^{(\pi)}+1$} represent the first time frame when {\small$x_{m,k}^{(\pi)}(t) > x_{m,k}^{(\pi)}(t+1)$}. And, we define {\small$t_{l,0}^{(\pi)} \triangleq$} min {\small$\left\{ t_{l\downarrow}^{(\pi)},t^{(\pi)}+\lceil r\rceil-1, t^{(\pi)}+K\right\}$}.

To prove (\ref{eq:tail_1}), we should prove 

{\small
\begin{equation*}
\begin{aligned}
	& \sum\limits_{m}\sum\limits_{k}b_{m,k}{\left[x_{m,k}^{(\pi)}(t^{(\pi)})-x_{m,k}^{(\pi)}(t^{(\pi)}-1)\right]}^{+}\\
  & \leq \eta \sum\limits_{t=t^{(\pi)}}^{t_{l,0}^{(\pi)}}\left[x_{m,k}^{(\pi)}(t)  +\frac{\varepsilon}{\mathcal{M}\mathcal{K}}\right]D^{(\pi)}(t)\\
	& + b_{m,k}\left[x_{m,k}^{(\pi)}(t^{(\pi)})+\frac{\varepsilon}{\mathcal{M}\mathcal{K}}\right]\ln\left(\frac{1+\frac{\varepsilon}{\mathcal{M}\mathcal{K}}}{x_{m,k}^{(\pi)}(t_{l,0}^{(\pi)})+\frac{\varepsilon}{\mathcal{M}\mathcal{K}}}\right)\\
	& -b_{m,k}\left[x_{m,k}^{(\pi)}(t^{(\pi)})+\frac{\varepsilon}{\mathcal{M}\mathcal{K}}\right]\ln\left(\frac{1+\frac{\varepsilon}{\mathcal{M}\mathcal{K}}}{x_{m,k}^{(\pi)}(t^{(\pi)})+\frac{\varepsilon}{\mathcal{M}\mathcal{K}}}\right)\\
	& \leq \eta \sum\limits_{t=t^{(\pi)}}^{t_{l,0}^{(\pi)}}\left[x_{m,k}^{(\pi)}(t)+\frac{\varepsilon}{\mathcal{M}\mathcal{K}}\right]D^{(\pi)}(t).
\end{aligned}
\end{equation*}}

(i) If {\small $x_{m,k}^{(\pi)}(t^{(\pi)})-x_{m,k}^{(\pi)}(t^{(\pi)}-1) \leq 0$}, then {\small${\left[x_{m,k}^{(\pi)}(t^{(\pi)})-x_{m,k}^{(\pi)}(t^{(\pi)}-1)\right]}^{+}=0$}. The inequality (\ref{eq:tail_1}) holds as the RHS of (\ref{eq:tail_1}) is greater than or equal to zero.

(ii) If {\small $x_{m,k}^{(\pi)}(t^{(\pi)})-x_{m,k}^{(\pi)}(t^{(\pi)}-1)> 0$}, then {\small $x_{m,k}^{(\pi)}(t^{(\pi)}) > x_{m,k}^{(\pi)}(t^{(\pi)}-1) $}. As {\small $x_{m,k}^{(\pi)}(t^{(\pi)}-1)>0$}, then {\small $x_{m,k}^{(\pi)}(t^{(\pi)})>0$}. Therefore, inequality {\small$t_{l,0}^{(\pi)}\leq t_{l\downarrow}^{(\pi)} $} holds. And inequality {\small $x_{m,k}^{(\pi)}(t)\geq x_{m,k}^{(\pi)}(t^{(\pi)}-1)$} holds when {\small$t \in [t^{(\pi)},t_{l,0}^{(\pi)}]$}. Accordingly, {\small $x_{m,k}^{(\pi)}(t) > 0$} holds for all {\small$t \in [t^{(\pi)},t_{l,0}^{(\pi)}]$}. According to the equation of Optimality condition, we have

\vspace{-2mm}
{\small
\begin{equation}
\label{eq:tail_1_Optimality}
\begin{aligned}
	 l_{m,k} -\sum_{n}  \theta_{m,n,k}^{(\pi)}(t) +r_k \rho_m^{(\pi)}(t)+\beta_{m,k}^{(\pi)}(t) -\beta_{m,k}^{(\pi)}(t+1)   = 0, \forall m,k,t \in [t^{(\pi)},t_{l,0}^{(\pi)}],
\end{aligned}
\end{equation}}	
\vspace{-2mm}

\noindent where {\small $\beta_{m,k}^{(\pi)}(t^{(\pi)})= \frac{b_{m,k}}{\eta}\ln(\frac{1+\frac{\varepsilon}{\mathcal{M}\mathcal{K}}}{x_{m,k}^{(\pi)}(t^{(\pi)}-1)+\frac{\varepsilon}{\mathcal{M}\mathcal{K}}})$}. According to (\ref{eq:tail_1_Optimality}) and the Complementary slackness conditions, we can conduct

\vspace{-3mm}
{\small
\begin{equation}
\label{eq:tail_1_2}
\begin{aligned}
  & \sum\limits_{t=t^{(\pi)}}^{t_{l,0}^{(\pi)}}l_{m,k}[x_{m,k}^{(\pi)}(t)+\frac{\varepsilon}{\mathcal{M}\mathcal{K}}] 
  + \sum\limits_{t=t^{(\pi)}}^{t_{l,0}^{(\pi)}}d_{m,n}{\lambda_{n,k}(t)}{y^{(\pi)}_{m,n,k}(t)}+\sum\limits_{t=t^{(\pi)}+1}^{t_{l,0}^{(\pi)}}b_{m,k}z^{(\pi)}_{m,k}(t)\\
  & +\frac{b_{m,k}}{\eta}\left[x^{(\pi)}_{m,k}(t^{(\pi)})+\frac{\varepsilon}{\mathcal{M}\mathcal{K}}\right]\ln\left(\frac{1+\frac{\varepsilon}{\mathcal{M}\mathcal{K}}}{x_{m,k}^{(\pi)}(t^{(\pi)}-1)+\frac{\varepsilon}{\mathcal{M}\mathcal{K}}}\right)\\
  & = \sum\limits_{t=t^{(\pi)}}^{t_{l,0}^{(\pi)}}l_{m,k}[x_{m,k}^{(\pi)}(t)+\frac{\varepsilon}{\mathcal{M}\mathcal{K}}] 
  + \sum\limits_{t=t^{(\pi)}}^{t_{l,0}^{(\pi)}}d_{m,n}{\lambda_{n,k}(t)}{y^{(\pi)}_{m,n,k}(t)}+\sum\limits_{t=t^{(\pi)}+1}^{t_{l,0}^{(\pi)}}b_{m,k}z^{(\pi)}_{m,k}(t)\\
  & +\frac{b_{m,k}}{\eta}\left[x^{(\pi)}_{m,k}(t^{(\pi)})+\frac{\varepsilon}{\mathcal{M}\mathcal{K}}\right]\ln\left(\frac{1+\frac{\varepsilon}{\mathcal{M}\mathcal{K}}}{x_{m,k}^{(\pi)}(t^{(\pi)}-1)+\frac{\varepsilon}{\mathcal{M}\mathcal{K}}}\right)\\
 & - \left[x_{m,k}^{(\pi)}(t^{(\pi)})+\frac{\varepsilon}{\mathcal{M}\mathcal{K}}\right]\left[l_{m,k} -\sum_{n} \theta_{m,n,k}^{(\pi)}(t)+r_k \rho_m^{(\pi)}(t)
 +\frac{b_{m,k}}{\eta}\ln\left(\frac{ 1 +\frac{\varepsilon}{\mathcal{M}\mathcal{K}}}{ x_{m,k}^{(\pi)}(t^{(\pi)}-1)+\frac{\varepsilon}{\mathcal{M}\mathcal{K}}}\right) -\beta_{m,k}^{(\pi)}(t^{(\pi)}+1) \right]\\ 
  & - \sum\limits_{t=t^{(\pi)}+1}^{t_{l,0}^{(\pi)}} \left[x_{m,k}^{(\pi)}(t)+\frac{\varepsilon}{\mathcal{M}\mathcal{K}}\right]\left[l_{m,k} -\sum_{n} \theta_{m,n,k}^{(\pi)}(t)
  +r_k\rho_m^{(\pi)}(t) + \beta_{m,k}^{(\pi)}(t) -\beta_{m,k}^{(\pi)}(t+1) \right]\\
  & - \sum\limits_{t=t^{(\pi)}+1}^{t_{l,0}^{(\pi)}} z_{m,k}^{(\pi)}(t) \left[\beta_{m,k}^{(\pi)}(t)-b_{m,k}\right]
  - \sum\limits_{t=t^{(\pi)}+1}^{t_{l,0}^{(\pi)}} \beta_{m,k}^{(\pi)}(t)\left[ x_{m,k}^{(\pi)}(t) - x_{m,k}^{(\pi)}(t-1) - z_{m,k}^{(\pi)}(t)\right]\\
  & - \sum\limits_{t=t^{(\pi)}}^{t_{l,0}^{(\pi)}} \mu_m^{(\pi)}(t)\left[\sum_{k}\sum_{n}c_k\lambda_{n,k}(t) y_{m,n,k}^{(\pi)}(t)-C_m\right].
\end{aligned}
\end{equation}}
\vspace{-6mm}

By rearranging the RHS of (\ref{eq:tail_1_2}), we have
\vspace{-2mm}
{\small
\begin{equation}
\label{eq:tail_1_3}
\begin{aligned}
 \sum\limits_{t=t^{(\pi)}}^{t_{l,0}^{(\pi)}}& l_{m,k}\left[x_{m,k}^{(\pi)}(t)+\frac{\varepsilon}{\mathcal{M}\mathcal{K}}\right]
+ \sum\limits_{t=t^{(\pi)}}^{t_{l,0}^{(\pi)}}d_{m,n}{\lambda_{n,k}(t)}{y^{(\pi)}_{m,n,k}(t)}+\sum\limits_{t=t^{(\pi)}+1}^{t_{l,0}^{(\pi)}}b_{m,k}z^{(\pi)}_{m,k}(t)\\
  & +\frac{b_{m,k}}{\eta}\left[x_{m,k}^{(\pi)}(t_m^{(\pi)})+\frac{\varepsilon}{\mathcal{M}\mathcal{K}}\right]\ln\left(\frac{1+\frac{\varepsilon}{\mathcal{M}\mathcal{K}}}{x_{m,k}^{(\pi)}(t^{(\pi)}-1)+\frac{\varepsilon}{\mathcal{M}\mathcal{K}}}\right)\\
  & = \sum\limits_{t=t^{(\pi)}}^{t_{l,0}^{(\pi)}}\left[ x_{m,k}^{(\pi)}(t)+\frac{\varepsilon}{\mathcal{M}\mathcal{K}}\right]D^{(\pi)}(t) + \left[x^{(\pi)}_{m,k}(t_{l,0}^{(\pi)})+\frac{\varepsilon}{\mathcal{M}\mathcal{K}}\right] \beta_{m,k}^{(\pi)}(t_{l,0}^{(\pi)}+1).
\end{aligned}
\end{equation}}
\noindent By rearranging the terms of (\ref{eq:tail_1_3}), we get 
\vspace{-2mm}
{\small
\begin{equation*}
\label{eq:tail_1_4}
\begin{aligned}
  & \frac{b_{m,k}}{\eta}\left[x^{(\pi)}_{m,k}(t^{(\pi)})+\frac{\varepsilon}{\mathcal{M}\mathcal{K}}\right]\ln\left(\frac{1+\frac{\varepsilon}{\mathcal{M}\mathcal{K}}}{x_{m,k}^{(\pi)}(t^{(\pi)}-1)+\frac{\varepsilon}{\mathcal{M}\mathcal{K}}}\right) \\
  & = \sum\limits_{t=t^{(\pi)}}^{t_{l,0}^{(\pi)}}\left[ x^{(\pi)}_{m,k}(t)+\frac{\varepsilon}{\mathcal{M}\mathcal{K}}\right]D^{(\pi)}(t)
  + \left[x^{(\pi)}_{m,k}(t_{l,0}^{(\pi)})+\frac{\varepsilon}{\mathcal{M}\mathcal{K}}\right] \beta_{m,k}^{(\pi)}(t_{l,0}^{(\pi)}+1)\\
  & -\sum\limits_{t=t^{(\pi)}}^{t_{l,0}^{(\pi)}}l_{m,k}\left[x_{m,k}^{(\pi)}(t)+\frac{\varepsilon}{\mathcal{M}\mathcal{K}}\right]
  - \sum\limits_{t=t^{(\pi)}}^{t_{l,0}^{(\pi)}}d_{m,n}{\lambda_{n,k}(t)}{y^{(\pi)}_{m,n,k}(t)}-\sum\limits_{t=t^{(\pi)}+1}^{t_{l,0}^{(\pi)}}b_{m,k}z^{(\pi)}_{m,k}(t).
\end{aligned}
\end{equation*}}
\vspace{-2mm}

Then, we confirm the final result by taking three cases into consideration, {\em i.e.,} {\small$t_{l,0}^{(\pi)}=t_{l\downarrow}^{(\pi)}$}, {\small$t_{l,0}^{(\pi)}=t^{(\pi)}+\lceil r\rceil-1$}, and  {\small$t_{l,0}^{(\pi)}=t^{(\pi)}+L$}. In the following, we only prove one case.

(a). If {\small$t^{(\pi)}+L\leq t_{l\downarrow}^{(\pi)}$} and {\small$t^{(\pi)}+L\leq t^{(\pi)}+\lceil r\rceil-1$}, we obtain {\small$t_{l,0}^{(\pi)}=t^{(\pi)}+L$}. And, based on the definition of
{\small $\beta_{m,k}^{(\pi)}(t^{(\pi)})$}, we derive 

\vspace{-2mm}
{\small
\begin{equation*}
\label{eq:tail_1_5}
\begin{aligned}
 \beta_{m,k}^{(\pi)}(t_{l,0}^{(\pi)}+1) = \frac{b_{m,k}}{\eta}\ln\left(\frac{1+\frac{\varepsilon}{\mathcal{M}\mathcal{K}}}{x_{m,k}^{(\pi)}(t_{l,0}^{(\pi)}-1)+\frac{\varepsilon}{\mathcal{M}\mathcal{K}}}\right).
\end{aligned}
\end{equation*}	}
\vspace{-2mm}
\noindent Accordingly, we get

{\small
\begin{equation*}
\label{eq:tail_1_6}
\begin{aligned}
	& \sum\limits_{m}\sum\limits_{k}b_{m,k}{\left[x_{m,k}^{(\pi)}(t^{(\pi)})-x_{m,k}^{(\pi)}(t^{(\pi)}-1)\right]}^{+}\\
	& \leq  \sum\limits_{m}\sum\limits_{k}b_{m,k}\left[x_{m,k}^{(\pi)}(t^{(\pi)})+\frac{\varepsilon}{\mathcal{M}\mathcal{K}}\right]\ln \left(\frac{{x_{m,k}^{(\pi)}(t^{(\pi)})}+\frac{\varepsilon}{\mathcal{M}\mathcal{K}}}{x_{m,k}^{(\pi)}(t^{(\pi)}-1)+\frac{\varepsilon}{\mathcal{M}\mathcal{K}}}\right)\\
	& = \sum\limits_{m}\sum\limits_{k} b_{m,k}\left[x_{m,k}^{(\pi)}(t^{(\pi)})+\frac{\varepsilon}{\mathcal{M}\mathcal{K}}\right]\ln\left(\frac{1+\frac{\varepsilon}{\mathcal{M}\mathcal{K}}}{x_{m,k}^{(\pi)}(t^{(\pi)}-1)+\frac{\varepsilon}{\mathcal{M}\mathcal{K}}}\right) \\
	&- \sum\limits_{m}\sum\limits_{k}b_{m,k}\left[x_{m,k}^{(\pi)}(t^{(\pi)})+\frac{\varepsilon}{\mathcal{M}\mathcal{K}}\right]\ln\left(\frac{1+\frac{\varepsilon}{\mathcal{M}\mathcal{K}}}{x_{m,k}^{(\pi)}(t^{(\pi)})+\frac{\varepsilon}{\mathcal{M}\mathcal{K}}}\right) \\
	& = \eta \bigg \{ \sum\limits_{t=t^{(\pi)}}^{t_{l,0}^{(\pi)}}\left[x_{m,k}^{(\pi)}(t)  +\frac{\varepsilon}{\mathcal{M}\mathcal{K}}\right]D^{(\pi)}(t) + \left[x_{m,k}^{(\pi)}(t_{l,0}^{(\pi)}) +\frac{\varepsilon}{\mathcal{M}\mathcal{K}}\right]\beta_{m,k}^{(\pi)}(t_{l,0}^{(\pi)}-1)\\
	& - \sum\limits_{t=t^{(\pi)}}^{t_{l,0}^{(\pi)}}l_{m,k}\left[x_{m,k}^{(\pi)}(t)+\frac{\varepsilon}{\mathcal{M}\mathcal{K}}\right]-  \sum\limits_{t=t^{(\pi)}}^{t_{l,0}^{(\pi)}}d_{m,n}{\lambda_{n,k}(t)}{y^{(\pi)}_{m,n,k}(t)}
	-\sum\limits_{t=t^{(\pi)}+1}^{t_{l,0}^{(\pi)}}b_{m,k}z^{(\pi)}_{m,k}(t)\bigg\}\\
	& - b_{m,k}\left[x_{m,k}^{(\pi)}(t)+\frac{\varepsilon}{\mathcal{M}\mathcal{K}}\right]\ln\left(\frac{1+\frac{\varepsilon}{\mathcal{M}\mathcal{K}}}{x_{m,k}^{(\pi)}(t^{(\pi)})+\frac{\varepsilon}{\mathcal{M}\mathcal{K}}}\right).
\end{aligned}
\end{equation*}}
\vspace{-6mm}

Finally, we can conduct 

{\small
\begin{equation*}
\label{eq:tail_1_7}
\begin{aligned}
	& \sum\limits_{m}\sum\limits_{k}b_{m,k}{\left[x_{m,k}^{(\pi)}(t^{(\pi)})-x_{m,k}^{(\pi)}(t^{(\pi)}-1)\right]}^{+}\\
	& \leq \eta \sum\limits_{t=t^{(\pi)}}^{t_{l,0}^{(\pi)}}\left[x_{m,k}^{(\pi)}(t)  +\frac{\varepsilon}{\mathcal{M}\mathcal{K}}\right]D^{(\pi)}(t) \\
    & + b_{m,k}\left[x_{m,k}^{(\pi)}(t^{(\pi)})+\frac{\varepsilon}{\mathcal{M}\mathcal{K}}\right]\ln\left(\frac{1+\frac{\varepsilon}{\mathcal{M}\mathcal{K}}}{x_{m,k}^{(\pi)}(t_{l,0}^{(\pi)})+\frac{\varepsilon}{\mathcal{M}\mathcal{K}}}\right)\\
    & - b_{m,k}\left[x_{m,k}^{(\pi)}(t^{(\pi)})+\frac{\varepsilon}{\mathcal{M}\mathcal{K}}\right]\ln\left(\frac{1+\frac{\varepsilon}{\mathcal{M}\mathcal{K}}}{x_{m,k}^{(\pi)}(t^{(\pi)})+\frac{\varepsilon}{\mathcal{M}\mathcal{K}}}\right)\\
	& \leq \eta \sum\limits_{t=t^{(\pi)}}^{t_{l,0}^{(\pi)}}\left[x_{m,k}^{(\pi)}(t)  +\frac{\varepsilon}{\mathcal{M}\mathcal{K}}\right]D^{(\pi)}(t) .
\end{aligned}
\end{equation*}}	 
According to the complementary slackness conditions, we can conduct inequality (\ref{eq:tail_1}) is true.

B. Proof of (\ref{eq:tail_2}).

 To prove (\ref{eq:tail_2}), we leverage the conclusion of the following lemma.

{\bf\em Lemma 7}. For each {\small$ORA^{(\pi)}$}, the sum of left-hand-side in (\ref{eq:tail_2}) can be bounded as shows:

\vspace{-2mm}
{\small
\begin{equation}
\label{eq:tail_2_1}
\begin{aligned}
	&\sum\limits_{m}\sum\limits_{k}{\psi}_{m,k}^{(\pi)}(t_b^{(\pi)}) +\sum\limits_{m}\sum\limits_{k}{\phi}_{m,k}^{(\pi)}(t_e^{(\pi)})  + \sum\limits_{0 \leq t_m^{(\pi)}\leq T-L-1 }\sum\limits_{m}\sum\limits_{k}\left[{\phi}_{m,k}^{(\pi)}(t_m^{(\pi)}) + {\psi}_{m,k}^{(\pi)}(t_m^{(\pi)})\right] \\
	&\leq \sum\limits_{ t^{(\pi)}\in \{t_b^{(\pi)},t_m^{(\pi)},t_e^{(\pi)}\} }\sum\limits_{m}\sum\limits_{k}\left[ x_{m,k}^{(\pi)}(t^{(\pi)}+L) -x_{m,k}^{(\pi)}(t^{(\pi)})   \right]  \cdot \frac{b_{m,k}}{\eta}\ln\left(\frac{1+\frac{\varepsilon}{\mathcal{M}\mathcal{K}}}{x_{m,k}^{(\pi)}(t^{(\pi)}-1)+\frac{\varepsilon}{\mathcal{M}\mathcal{K}}}\right).
\end{aligned}
\end{equation}}	
\vspace{-2mm}

\noindent Let {\small$ {\phi}_{m,k}^{(\pi)}(t_m^{(\pi)})$} denote the RHS of (\ref{eq:tail_2_1}), {\em i.e.,}

\vspace{-3mm}
{\small
\begin{equation*}
\label{eq:tail_2_2}
\begin{aligned}
	 \Phi_{m,k}^{(\pi)}(t^{(\pi)})=  \left[ x_{m,k}^{(\pi)}(t^{(\pi)}+L) -x_{m,k}^{(\pi)}(t^{(\pi)})   \right]  \cdot \frac{b_{m,k}}{\eta}\ln\left[\frac{1+\frac{\varepsilon}{\mathcal{M}\mathcal{K}}}{x_{m,k}^{(\pi)}(t^{(\pi)}-1)+\frac{\varepsilon}{\mathcal{M}\mathcal{K}}}\right].
\end{aligned}
\end{equation*}}	
\vspace{-3mm}

\noindent Let {\small$t_{l\uparrow}^{(\pi)}+1$} represent the last time frame when {\small$x_{m,k}^{(\pi)}(t)$} increases. Furthermore, we define {\small$t_{l,1}^{(\pi)} =$} max {\small$\{ t_{l\uparrow}^{(\pi)}, t^{(\pi)}+L-\lceil r\rceil+1 \}$}. To prove (\ref{eq:tail_2}), we should prove that, at each {\small$t^{(\pi)}$},

\vspace{-2mm}
{\small
\begin{equation}
\label{eq:tail_2_3}
\begin{aligned}
  \Phi_{m,k}^{(\pi)}(t^{(\pi)}) \leq &  \eta \sum\limits_{t=t_{l,1}^{(\pi)}}^{t^{(\pi)}+L}	\left [x_{m,k}^{(\pi)}(t)+\frac{\varepsilon}{\mathcal{M}\mathcal{K}}\right]D^{(\pi)}(t).
\end{aligned}
\end{equation}}	
\vspace{-2mm}

(i) If {\small $x_{m,k}^{(\pi)}(t^{(\pi)}+L)-x_{m,k}^{(\pi)}(t^{(\pi)}) \leq 0$}, then {\small$\Phi_{m,k}^{(\pi)}(t^{(\pi)}) \leq 0$}. The inequality (\ref{eq:tail_2}) holds as the RHS of (\ref{eq:tail_2}) is greater than or equal to zero.

(ii) If {\small $x_{m,k}^{(\pi)}(t^{(\pi)}+L)-x_{m,k}^{(\pi)}(t^{(\pi)})> 0$}, then {\small $x_{m,k}^{(\pi)}(t^{(\pi)}+L) > x_{m,k}^{(\pi)}(t^{(\pi)}) $}. As {\small $x_{m,k}^{(\pi)}(t^{(\pi)})\geq 0$}, then {\small $x_{m,k}^{(\pi)}(t^{(\pi)}+L)>0$}. Thus, 
\vspace{-2mm}
{\small
\begin{equation}
\label{eq:tail_2_4}
\begin{aligned}
	 \Phi_{m,k}^{(\pi)}(t^{(\pi)})  \leq    b_{m,k} \left[ x_{m,k}^{(\pi)}(t^{(\pi)}+L) -x_{m,k}^{(\pi)}(t^{(\pi)})   \right] \leq  b_{m,k}\cdot x_{m,k}^{(\pi)}(t^{(\pi)}+L).
\end{aligned}
\end{equation}}	
\vspace{-2mm}

Furthermore, as {\small$t_{l,1}^{(\pi)}\geq t_{l\uparrow}^{(\pi)} $}, the inequality {\small $x_{m,k}^{(\pi)}(t)\geq x_{m,k}^{(\pi)}(t^{(\pi)}+L)$} holds for all {\small$t \in [t_{l,1}^{(\pi)},t^{(\pi)}+L]$}. Thus, the inequality {\small $x_{m,k}^{(\pi)}(t)>0$} holds for all {\small$t \in [t_{l,1}^{(\pi)},t^{(\pi)}+L]$}. According to the optimality condition, we get

\vspace{-2mm}
{\small
\begin{equation}
\label{eq:tail_2_Optimality}
\begin{aligned}
	 l_{m,k} -\sum_{n} \theta_{m,n,k}^{(\pi)}(t) +  r_k \rho_m^{(\pi)}(t)+\beta_{m,k}^{(\pi)}(t)  -\beta_{m,k}^{(\pi)}(t+1)   = 0,  \forall t \in [t_{l,1}^{(\pi)},t^{(\pi)}+L],
\end{aligned}
\end{equation}}	
\vspace{-3mm}

\noindent where {\small $\beta_{m,k}^{(\pi)}(t^{(\pi)}+L+1)= \frac{b_{m,k}}{\eta}\ln(\frac{1+\frac{\varepsilon}{\mathcal{M}\mathcal{K}}}{x_{m,k}^{(\pi)}(t^{(\pi)}+L)+\frac{\varepsilon}{\mathcal{M}\mathcal{K}}})$}. 
By adding (\ref{eq:tail_2_Optimality}) from time {\small$ t_{l,1}^{(\pi)}$} to {\small$t^{(\pi)}+L$}, we get

\vspace{-3mm}
{\small
\begin{equation}
\label{eq:tail_2_Optimality_1}
\begin{aligned}
	 \sum\limits_{t=t_{l,1}^{(\pi)}}^{t^{(\pi)}+L}\sum_{n} \theta_{m,n,k}^{(\pi)}(t)  =  l_{m,k} + \sum\limits_{t=t_{l,1}^{(\pi)}}^{t^{(\pi)}+L} r_k \rho_m^{(\pi)}(t)+\beta_{m,k}^{(\pi)}(t_{l,1}^{(\pi)})    -\frac{b_{m,k}}{\eta}\ln\left(\frac{1+\frac{\varepsilon}{\mathcal{M}\mathcal{K}}}{x_{m,k}^{(\pi)}(t^{(\pi)}+L)+\frac{\varepsilon}{\mathcal{M}\mathcal{K}}}\right).
\end{aligned}
\end{equation}}	
\vspace{-3mm}

\noindent Based on (\ref{eq:tail_2_Optimality_1}), we obtain

\vspace{-3mm}
{\small
\begin{equation}
\label{eq:tail_2_5}
\begin{aligned}
  \eta \sum\limits_{t=t_{l,1}^{(\pi)}}^{t^{(\pi)}+L} \left[x_{m,k}^{(\pi)}(t)  +\frac{\varepsilon}{\mathcal{M}\mathcal{K}}\right] D^{(\pi)}(t) \geq \eta \left[x_{m,k}^{(\pi)}(t^{(\pi)}+L)  +\frac{\varepsilon}{\mathcal{M}\mathcal{K}}\right] \sum\limits_{t=t_{l,1}^{(\pi)}}^{t^{(\pi)}+L} D^{(\pi)}(t) \geq b_{m,k}x_{m,k}^{(\pi)}(t^{(\pi)}+L). 
\end{aligned}
\end{equation}}	
\vspace{-3mm}

Therefore, (\ref{eq:tail_2_4}) and (\ref{eq:tail_2_5}) together indicate that (\ref{eq:tail_2_3}) is true. According to the Complementary slackness conditions, we can conduct (\ref{eq:tail_2}) is true.\qed

\vspace{2mm} 
{\bf Appendix D }

{\bf\em Important conclusions from Lemma 3.} 

Firstly, from (\ref{eq:tail_1}), we have the following conclusions.

 When {\small$\lceil r\rceil < L+1 $}, then {\small$t_{l,0}^{(\pi)}=$}  min {\small$\left\{ t_{l\downarrow}^{(\pi)},t^{(\pi)}+\lceil r\rceil-1\right\}$}. Therefore, 

(i) If the episodes’ beginning time {\small$t^{(\pi)} \in \left[1,T-\lceil r\rceil\right]$}, based on (\ref{eq:tail_1}) and {\small$t_{l,0}^{(\pi)} \leq T-\lceil r\rceil-1$}, we get

\vspace{-3mm}
{\small
\begin{equation*}
\label{eq:ORA_16}
\begin{aligned}
  \sum\limits_{m}\sum\limits_{k}{\Omega}_{m,k}^{(\pi)}(t^{(\pi)})\leq \eta \sum\limits_{t=t^{(\pi)}}^{t^{(\pi)}+\lceil r\rceil-1}\left[x_{m,k}^{(\pi)}(t) +\frac{\varepsilon}{\mathcal{M}\mathcal{K}}\right]D^{(\pi)}(t) \leq\eta(1+\frac{\varepsilon}{\mathcal{M}\mathcal{K}})\sum\limits_{t=t^{(\pi)}}^{t^{(\pi)}+\lceil r\rceil-1}D^{(\pi)}(t).
\end{aligned}
\end{equation*}}
\vspace{-3mm}

\noindent The last inequality can follow from the complementary slackness condition.

(ii) If the episodes’ beginning time {\small$t^{(\pi)} \in [T-\lceil r\rceil+1,T]$}, we have that {\small$t_{l,0}^{(\pi)} \leq T$} according to {\small$t_{l\downarrow}^{(\pi)} \leq T$}. Based on (\ref{eq:tail_1}), we conduct 

\vspace{-2mm}
{\small
\begin{equation*}
\label{eq:ORA_17}
\begin{aligned}
  \sum\limits_{m}\sum\limits_{k}{\Omega}_{m,k}^{(\pi)}(t^{(\pi)})\leq \eta \sum\limits_{t=t^{(\pi)}}^{T}\left[x_{m,k}^{(\pi)}(t) +\frac{\varepsilon}{\mathcal{M}\mathcal{K}}\right]D^{(\pi)}(t) \leq\eta(1+\frac{\varepsilon}{\mathcal{M}\mathcal{K}})\sum\limits_{t=t^{(\pi)}}^{T}D^{(\pi)}(t).
\end{aligned}
\end{equation*}}
\vspace{-2mm}

Secondly, from (\ref{eq:tail_2}), we conduct the following conclusions.

When {\small$\lceil r\rceil < L+1 $}, then {\small$t_{l,1}^{(\pi)} =\max \left\{ t_{l\uparrow}^{(\pi)},t^{(\pi)}+L-\lceil r\rceil+1\right\}$}. Therefore, 

(i) If the episodes’ beginning time {\small$t^{(\pi)} \in [-L+\lceil r\rceil+1,T-L]$}, according to (\ref{eq:tail_2}) and {\small$t_{l,1}^{(\pi)} \geq t^{(\pi)}+L-\lceil r\rceil+1$}, we get 

\vspace{-2mm}
{\small
\begin{equation*}
\label{eq:ORA_19}
\begin{aligned}
  \sum\limits_{m}\sum\limits_{k}{\phi}_{m,k}^{(\pi)}(t^{(\pi)})\leq \eta \sum\limits_{t=t^{(\pi)}+L\atop-\lceil r\rceil+1}^{t^{(\pi)}+L}\left[x_{m,k}^{(\pi)}(t) +\frac{\varepsilon}{\mathcal{M}\mathcal{K}}\right]D^{(\pi)}(t)  \leq\eta(1+\frac{\varepsilon}{\mathcal{M}\mathcal{K}}) \sum\limits_{t=t^{(\pi)}+L\atop-\lceil r\rceil+1}^{t^{(\pi)}+L}D^{(\pi)}(t). 
\end{aligned}
\end{equation*}}

(ii) If the episodes’ beginning time {\small$t^{(\pi)} \in [-L+1,-L+\lceil r\rceil]$}, we obtain 
{\small$t_{l,1}^{(\pi)} \geq 1$} according to {\small$t_{l\uparrow}^{(\pi)} \geq1$}. Based on (\ref{eq:tail_2}), we derive 

{\small
\begin{equation*}
\label{eq:ORA_20}
\begin{aligned}
  \sum\limits_{m}\sum\limits_{k}{\phi}_{m,k}^{(\pi)}(t^{(\pi)})\leq \eta \sum\limits_{t=1}^{t^{(\pi)}+L}\left[x_{m,k}^{(\pi)}(t) +\frac{\varepsilon}{\mathcal{M}\mathcal{K}}\right]D^{(\pi)}(t) \leq\eta(1+\frac{\varepsilon}{\mathcal{M}\mathcal{K}})\sum\limits_{t=1}^{t^{(\pi)}+L}D^{(\pi)}(t). 
\end{aligned}
\end{equation*}}
\qed

{\bf Appendix E }

The mainly computational burden of algorithm $ORA$ is the interior point method to solve
 {\small $P_{ORA}^{(\pi)}$}. For algorithm $RDSP$, in each $while$ loop in the outermost: line $4-9$ iterate $O(KM)$ steps to terminate. Algorithm $RDSP$ runs $O(K^2M^2)$ steps. Hence, our proposed algorithm can be completed in a polynominal running time.  \qed

\end{document}